# Theory and phase-field simulations on electrical control of spin cycloids in a multiferroic

Fei Xue, Tiannan Yang, and Long-Qing Chen

Department of Materials Science and Engineering, The Pennsylvania State University, University Park, Pennsylvania 16802, USA

Cycloidal spin orders are common in multiferroics. One of the prototypical examples is $BiFeO_3$ (BFO) which shows a large polarization and a cycloidal antiferromagnetic order at room temperature. Here we employ Landau theory and phase-field simulations to analyze the coupled switching dynamics of polarization and cycloidal antiferromagnetic orders in BFO. We are able to identify 14 types of transitional spin structures between two cycloids and 9 electric-field-induced spin switching paths. We demonstrate the electric-field-induced rotation of wave vectors of the cycloidal spins and discover 2 types of cycloidal spin switching dynamics: fast local spin flips and slow rotation of wave vectors. Also, we construct road maps to achieve the switching between any two spin cycloids through multi-step applications of electric fields. The work provides a theoretical framework for the phenomenological description of spin cycloids and a fundamental understanding of the switching mechanisms to achieve electrical control of magnetic orders.

## 1. Introduction

In ferroelectrics and ferromagnets, electric and magnetic fields can be employed to manipulate their electric polarization and magnetization, respectively. However, the cross-control, namely the electrical control of magnetic order or magnetic control of electric order is highly nontrivial, and its fundamental understanding is critical to realizing next generation electronic devices [1,2]. Multiferroics possessing more than one ferroic orders are expected to exhibit strong magnetoelectric effects if its ferroelectric and magnetic orders are directly coupled to each other. In a class of multiferroics, ferroelectric polarization coexists with spin density waves, where the spin helicity is coupled to the polarization direction [3,4]. In these materials, it is found that the polarization can be switched by magnetic fields [5], and the spin helicity is strongly coupled to the direction of polarization [6].

$BiFeO_3$ (BFO) is a room-temperature multiferroic exhibiting a large spontaneous polarization of ~100 $\mu C/cm^2$ and a cycloidal antiferromagnetic (AFM) order [7]. It exists in the *R3c* perovskite structure with its polarization along one of its eight $\{111\}_{pseudocubic}$ directions [8]. Below the Neel temperature 370 $^oC$, the short-range magnetic ordering of BFO is G-type antiferromagnetic with each $Fe^{+3}$ spin surrounded by six antiparallel spins on the nearest Fe neighbors.

The spins on the nearest Fe neighbors are not exactly antiparallel to each other due to a weak canting moment caused by the local magnetoelectric coupling [9], and a net magnetization exists locally [10]. A long-range superstructure with a period of 62-64 nm is superimposed on this canted AFM order, resulting in an incommensurate spin cycloid of the AFM ordered sublattices [11]. The AFM easy plane, within which the spins rotate, is spanned by the polarization vector and a wave vector. The wave vector is perpendicular to polarization and along one of the 12 $\{1\bar{1}0\}_{pseudocubic}$ directions (Fig. 1a). The cycloidal spin order is known to exist in BFO bulk crystals [11,12] and was thought to be suppressed by epitaxial strains in BFO films [13,14]. Recently, the spin cycloid is also observed in BFO films [10,15-17], and it is possible to control the cycloidal spins using electric fields [10,16]. However, the cycloidal spin structures and electric-field-controlled spin switching dynamics are not well understood.

In this article, we combine Landau theory and phase-field simulations to predict and analyze the spin configurations and magnetoelectric switching dynamics of the cycloidal spin orders in BFO. A region with uniform polarization and spin wave vector is regarded as one AFM domain. We identify 24 AFM domains and 14 types of domain walls and calculate the corresponding magnetic contributions to the domain wall energies. It is shown that there are 9 types of possible cycloidal spin switching paths following the switching of polarization by an electric field, where the wave vector is rotated by 0°, 60°, 90°, 120°, or 180°. When the wave vector maintains its direction or is rotated by 180° during the whole process, the switching is fast and only involves the local flips of spins. For the other cases, the switching is characterized by two steps: local spin flips while maintaining the wave vector and growth of the domains with a new wave vector. Furthermore, road maps are constructed to achieve the transition between any two AFM domains through multi-step applications of electric fields.

## 2. Theory, results, and discussion

### 2.1 Properties of a general spin cycloid

A spin cycloid is a sequence of spins which can be treated as the trace of a vector on a circle when the circle rolls along another vector (wave vector), as illustrated in Figs. 1(a) and 1(b). Thus the spins in a cycloid are rotated within a 2D plane. If we define the direction of the rotation axis based on the right-hand rule, a spin cycloid can be described by two vectors, the wave vector **k** and the vector for the rotation axis **m**. The spins are essentially static, and a cycloid can be described by one of two antiparallel wave vectors (Fig. 1a). Note that when the wave vector **k** is switched to its opposite direction, the rotation axis **m** is also flipped to its opposite direction. If we label the cycloid in Fig. 1(a) by $C(\mathbf{k},\mathbf{m})$, it can also be labelled by $C(-\mathbf{k},-\mathbf{m})$, i.e., $C(\mathbf{k},\mathbf{m}) = C(-\mathbf{k},-\mathbf{m})$. On the other hand, the cycloid with the other combinations of $\pm\mathbf{k}$ and $\pm\mathbf{m}$, i.e., $C(-\mathbf{k},\mathbf{m})$ and $C(\mathbf{k},-\mathbf{m})$, possesses the opposite rotation sense, which is shown in Fig. 1(b).

### 2.2 Description of spin cycloids in BFO

In this study, we neglect the small net ferromagnetic moment in BFO and assume that locally BFO is a G-type antiferromagnet, i.e., the nearest Fe neighbors possess antiparallel magnetic moments. In this case, the cycloid in BFO consists of the vectors representing the AFM order parameter, which is defined as $L_i = \frac{1}{2}(M_{ai} - M_{bi})(i = 1-3)$ and describes the staggered sublattice magnetization (Fig. 1c). Polarization **P** is coupled to the AFM order **L** through the coupling term $\gamma \mathbf{P} \cdot [\mathbf{L}(\nabla \cdot \mathbf{L}) - (\mathbf{L} \cdot \nabla)\mathbf{L}]$, which guarantees that **P** is perpendicular to both **k** and **m**. The coupling term is called the Lifshitz invariant, which is responsible for the appearance of the incommensurate spin cycloid [3,18]. The coupling term is also responsible for the opposite rotation senses of the spin cycloids associated with the antiparallel **P** vectors, as illustrated in Figs. 1(a) and 1(b).

Although 6 **k** vectors (along $\{110\}$) are perpendicular to a **P** vector (along $\{111\}$), the 6 **k** vectors correspond to 3 distinct spin cycloids, since antiparallel **k** vectors correspond to the same cycloid as shown in Fig. 1(a). Since there are 8 distinct **P** directions and each **P** direction corresponds to 3 distinct **k** directions, there exist a total of 24 variants of spin cycloids. We define the region with uniform **P** and **k** vectors as a domain. These 24 domains are symmetry-related and energy-degenerate. Since polarization is easily measured and controlled in BFO, instead of C(**k**, **m**), we can also use **P** and **k** vectors to label a cycloid, i.e., C(**P**, **k**). Due to the property of a cycloid as discussed in Section 2.1, in principle we can label a cycloid by either $C(\mathbf{P},\mathbf{k})$ or $C(\mathbf{P},-\mathbf{k})$, i.e., there exist two options of labels for each spin cycloid in BFO.

To give a unique label to each of the 24 spin cycloids, we choose three out of six **k** vectors for each polarization vector based on the following three rules. *Rule 1*, the **k** vector directions for $\mathbf{P} = [111]$ is chosen as $\mathbf{k}_1^0 = [1\bar{1}0]$, $\mathbf{k}_2^0 = [\bar{1}01]$, and $\mathbf{k}_3^0 = [01\bar{1}]$. As shown in Fig. 1(d), the angles between any two of the three **k** vectors are 120°. *Rule 2*, after **P** is switched by 71° from $[111]$ to $\mathbf{P}'$, one **k′** vector diection is chosen as $\mathbf{k}_1^0$, $\mathbf{k}_2^0$, or $\mathbf{k}_3^0$ (the mutual perpendicular relation between $\mathbf{P}'$ and **k′** should be maintained), and the other two **k′** vectors are assigned such that they form an angle of 120° with the first one. For example, if $\mathbf{P}' = [11\bar{1}]$, the first **k′** vector is $\mathbf{k}_1^0 = [1\bar{1}0]$, and the other two **k′** vectors are $[\bar{1}0\bar{1}]$, and $[011]$. Since three 71° switchings are possible for $\mathbf{P} = [111]$, based on the above two rules, the **k** vectors for four **P** vectors are determined. The remaining four **P** vectors can be obtained by 180° switching of the four **P** vectors that have already been assigned **k** vectors. *Rule 3*, if two **P** vectors are antiparallel, the sets of their **k** vectors also form antiparallel pairs. For example, for $\mathbf{P}=[\bar{1}\bar{1}\bar{1}]$, the **k** vectors are $-\mathbf{k}_1^0=[\bar{1}10]$, $-\mathbf{k}_2^0=[10\bar{1}]$, and $\mathbf{k}_3^0=-[0\bar{1}1]$. The three rules are designed for a simple and constant domain wall description, which will be discussed in Section 2.4. Based on these three rules, the labels for the 24 spin cycloids in BFO are listed in Table I.

Table I. List of 24 spin cycloids in BFO. A cyloid is labelled by $C(\mathbf{P},\mathbf{k})$, and each **P** vector corresponds to three **k** vectors.

| **P** | **k** |
|---|---|
| $[111]$ | $[1\bar{1}0]$, $[\bar{1}01]$, $[01\bar{1}]$ |
| $[\bar{1}\bar{1}\bar{1}]$ | $[\bar{1}10]$, $[10\bar{1}]$, $[0\bar{1}1]$ |
| $[11\bar{1}]$ | $[1\bar{1}0]$, $[\bar{1}0\bar{1}]$, $[011]$ |
| $[\bar{1}\bar{1}1]$ | $[\bar{1}10]$, $[101]$, $[0\bar{1}\bar{1}]$ |
| $[1\bar{1}1]$ | $[110]$, $[\bar{1}01]$, $[0\bar{1}\bar{1}]$ |
| $[\bar{1}1\bar{1}]$ | $[\bar{1}\bar{1}0]$, $[10\bar{1}]$, $[011]$ |
| $[\bar{1}11]$ | $[\bar{1}\bar{1}0]$, $[101]$, $[01\bar{1}]$ |
| $[1\bar{1}\bar{1}]$ | $[110]$, $[\bar{1}0\bar{1}]$, $[0\bar{1}1]$ |

## 2.3. Phase-field description of spin cycloids in BFO

Thermodynamics of the BFO system is described by three sets of order parameters, i.e., polarization $P_i(i=1-3)$, oxygen octahedral tilt $\theta_i(i=1-3)$ [19], and AFM order parameter $L_i(i=1-3)$. The total free energy of an inhomogeneous system containing polarization distribution, oxygen octahedral tilt, and AFM order is given by [15,19,20]

$$F = \int dV\{\alpha_{ij}P_iP_j + \alpha_{ijkl}P_iP_jP_kP_l + \beta_{ij}\theta_i\theta_j + \beta_{ijkl}\theta_i\theta_j\theta_k\theta_l + t_{ijkl}P_iP_j\theta_k\theta_l$$
$$+ K_1(L_1^2L_2^2 + L_1^2L_3^2 + L_2^2L_3^2) + K_2L_1^2L_2^2L_3^2 + h(\mathbf{L}\cdot\mathbf{P})^2 + \frac{1}{2}g_{ijkl}\frac{\partial P_i}{\partial x_j}\frac{\partial P_k}{\partial x_l} \quad , \tag{1}$$
$$+ \frac{1}{2}\kappa_{ijkl}\frac{\partial \theta_i}{\partial x_j}\frac{\partial \theta_k}{\partial x_l} + A\sum_{i=1,2,3}(\nabla L_i)^2 + \gamma\mathbf{P}\cdot[\mathbf{L}(\nabla\cdot\mathbf{L}) - (\mathbf{L}\cdot\nabla)\mathbf{L}]\}$$

where $\alpha_{ij}, \alpha_{ijkl}, \beta_{ij}, \beta_{ijkl}$ , and $t_{ijkl}$ are the coefficients for **P** and **θ** related to the dielectric and rotational susceptibilities and degree of coupling between **P** and **θ**; $K_1$ and $K_2$ are the anisotropy constants for **L**; $h$ is the biquadratic coupling coefficient between **L** and **P**; $g_{ijkl}$ and $\kappa_{ijkl}$ are the gradient energy coefficients for **P** and **θ**; *A* is the AFM exchange constant; $\gamma$ is the coefficient for the inhomogeneous magnetoelectric interaction [15]. Note that **L** is assumed to be only coupled to **P**, and independent of **θ**. The reason that **θ** is incorporated in Eq. (1) is to stabilize the Ising-like 180° ferroelectric domain walls [19,21]. The coefficients related to **P** and **θ** are taken from Ref. [19], and the values of *h*, *A*, and $\gamma$ are from Ref. [15]. The anisotropy constants are approximated to be $K_1 = -1.0\times10^4\ \mathrm{J\cdot m^{-3}}$, and $K_2 = 3.1\times10^4\ \mathrm{J\cdot m^{-3}}$, which produce an easy direction of **L** along $\{110\}_{pseudocubic}$ [22]. The complete form of Eq. (1) is provided in Section I of Supplementary Materials, and all the coefficients are listed in Table SI [23]. The coordinate system is rotated when needed [19].

The evolution of the **P** and **θ** vectors is described by the time-dependent Ginzburg-Landau (TDGL) equations

$$\frac{\partial P_i}{\partial t} = -K_P\frac{\delta F}{\delta P_i}, \frac{\partial \theta_i}{\partial t} = -K_\theta\frac{\delta F}{\delta \theta_i}, \tag{2}$$

where $K_P$ and $K_\theta$ are the kinetic coefficients related to the domain wall mobility. The TDGL equations are solved based on a semi-implicit spectral method [19,24-26]. Following Ref. [27], $K_P = K_\theta = 0.05$ *arb. unit* is assumed in solving Eq. (2).

The evolution of **L** at a fine time scale is complex as demonstrated in Refs. [28,29]. However, since the magnetic susceptibility of BFO is small at room temperature [30], at a time scale coarser than the resolution of 1 ps, the evolution of **L** can be approximated by a TDGL-type equation (for the details see Section III of Supplementary Materials [23])

$$\frac{\partial L_i}{\partial t} = \frac{\omega}{\upsilon}\mathbf{f}_{\mathrm{L}}, \tag{3}$$

where $\omega$ is the electron gyromagnetic ratio and $\nu$ the Gilbert damping constant. $\mathbf{f}_{\mathrm{L}}$ is the effective field acting on **L**, calculated as $\mathbf{f}_{\mathrm{L}} = -\frac{1}{\mu_0 L_S}\frac{\delta F}{\delta \mathbf{L}}$, where $\mu_0$ is the vacuum permeability, and $L_S$ is the saturation AFE magnetization ( $L_S \sim 5.6\times10^5\ \mathrm{A\cdot m^{-1}}$ following [31]). $\omega = 2.21\times10^5\ \mathrm{m\cdot A^{-1}\cdot s^{-1}}$ and $\upsilon = 0.5$, following Ref. [32]. Note that the order parameter **L** in Eq. (3) maintains its magnitude and only rotates its direction, different from the order parameters described by Eq. (2) which can change both their magnitude and directions. By solving Eq. (3), we simulate the spin relaxation behavior in the ps-ns regimes, neglecting the rapid oscillations of **L** in the sub-ps regime [28]. This is a reasonable approximation since we focus on the dynamics of spin cycloids with the switching time ranging from 20 ps to 10,000 ps, as will be shown later.

The phase-field model is employed to calculate the single-domain state of **L**. Given polarization **P** and wave vector **k**, the cycloidal distribution of **L** is given by [23]

$$\mathbf{L} = -\frac{\gamma\mathbf{P}}{|\gamma\mathbf{P}|}\cos(\mathbf{k}\cdot\mathbf{x}) + \frac{\mathbf{k}}{|\mathbf{k}|}\sin(\mathbf{k}\cdot\mathbf{x}), \tag{4}$$

where **x** is a spatial vector. In Eq. (4), the helicity of a cycloid is determined by the sign of $\gamma\mathbf{P}$ and independent of the sign of **k**, i.e., **k** and –**k** correspond to the same spin cycloid. This is consistent with the conclusions in Section 2.1.

The Lifshitz invariant guarantees that the system obtains its energy minima when **k** is perpendicular to **P** [3]. With **k** rotating within the plane perpendicular to **P**, the free energy versus the direction of **k** is plotted in Fig. 1(d), which shows six energy minima. Since two antiparallel **k** vectors correspond to the same spin cycloid, the six energy minima degenerate to three distinct distributions of **L**. The orientation anisotropy in Fig. 1(d) is caused by the anisotropy term (see Fig. S2b for the isotropic energy surface with $K_1 = K_2 = 0.0\ \mathrm{J\cdot m^{-3}}$ [23]). The system energy also depends on the wavelength of the spin cycloid, and the corresponding trend is plotted in Fig. S1(b) [23], which shows an equilibrium wavelength of ~62 nm. The equilibrium wavelength is consistent with experimental measurements [12]. In this paper, we assume that the cycloid wavelength maintains the equilibrium value, and we focus on the directions of the **P** and **k** vectors of the cycloids in BFO.

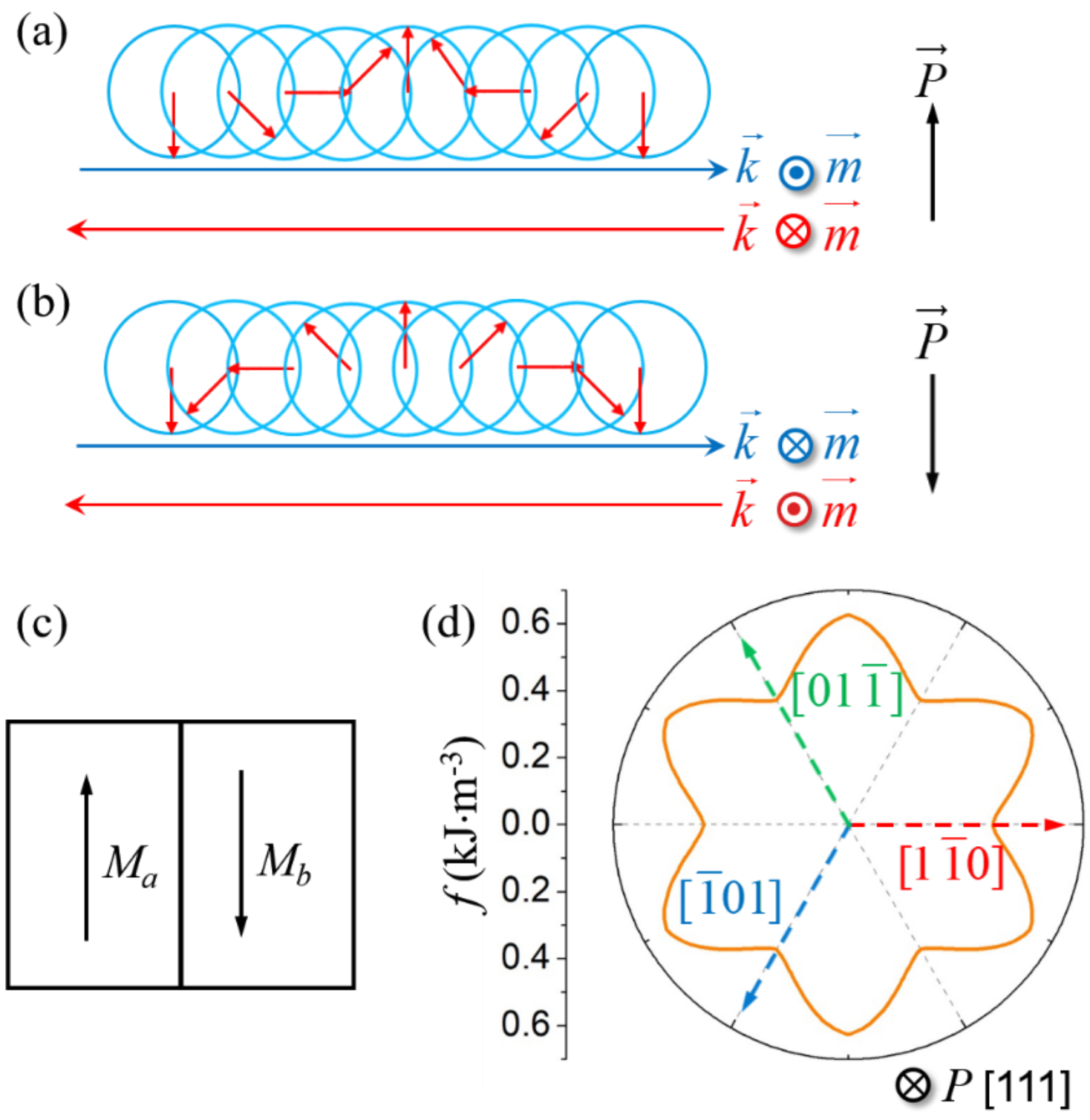

FIG. 1. Schematics and simulation results for AFM spin cycloids. (a) and (b) Spin cycloids with opposite rotation senses. The red arrows on the blue circles represent spin vectors. The black arrows denote polarization vectors. (c) Schematic for AFM sublattices. (d) Polar plot of the calculated free energy versus the direction of **k**.

### 2.4. Domain walls between spin cycloids in BFO

In this section we consider AFM domain walls, which are the transition regions between 2 spin cycloids. Similar to the classification of traditional ferroelectric domain walls [33], symmetry-related AFM domain walls are considered to belong to the same type. We need three quantities to describe a domain wall: the angle change of the **P** vector across a wall $\Delta\theta_{\mathbf{P}}$, the angle change of the **k** vector across a wall $\Delta\theta_{\mathbf{k}}$, and the domain wall orientation. We label a domain wall using the notation $W(\Delta\theta_{\mathbf{P}},\Delta\theta_{\mathbf{k}})$, and add more description for the domain wall orientation when a specific $W(\Delta\theta_{\mathbf{P}},\Delta\theta_{\mathbf{k}})$ domain wall corresponds to more than one wall orientations. $\Delta\theta_{\mathbf{P}}$ has four possible values: 0°, 71°, 109°, and 180°, while $\Delta\theta_{\mathbf{k}}$ has five options: 0°, 60°, 90°, 120°, and 180° [33]. Based on the assigned labels of spin cycloids listed in Table I, we calculate all the possible values of $\Delta\theta_{\mathbf{P}}$ and $\Delta\theta_{\mathbf{k}}$ by exhaustion, which are listed in Table II. Table II also shows the angle change of the **m** vector across the domain wall $\Delta\theta_{\mathbf{m}}$. Note that if two domain walls satisfy $\Delta\theta_{\mathbf{P}1}+\Delta\theta_{\mathbf{P}2}=180°$ and $\Delta\theta_{\mathbf{k}1}+\Delta\theta_{\mathbf{k}2}=180°$, then $\Delta\theta_{\mathbf{m}1}=\Delta\theta_{\mathbf{m}2}$. This is because the **m** vector can be obtained by the cross product of the **P** and **k** vectors.

In this study, we only consider the stress-free and charge-neutral domain walls with their orientations satisfying the mechanical compatibility condition and electrical neutrality condition [33]. Therefore, the elastic energy and electrostatic energy are neglected in Eq. (1). The phase-field simulations of the BFO domain walls with the elastic energy and electrostatic energy included can be found in Ref. [19]. Based on the value of $\Delta\theta_{\mathbf{P}}$, we discuss different domain walls in detail below.

The domain walls with $\Delta\theta_{\mathbf{P}}$=180° and $\Delta\theta_{\mathbf{P}}$=0° have more than one possible orientations. When $\Delta\theta_{\mathbf{P}}$=180°, due to the electrical neutrality condition, the domain wall normal **n** is confined to be perpendicular to **P** [33]. If we only consider domain walls with low Miller indices, **n** has the same set of 3 possible directions as **k**. For example, for two domains with $\mathbf{P}_1=[111]$ and $\mathbf{P}_2=[\bar{1}\bar{1}\bar{1}]$, the possible directions of **n** and **k** are both $\pm[1\bar{1}0]$, $\pm[\bar{1}01]$, and $\pm[01\bar{1}]$. As shown in Table II, $\Delta\theta_{\mathbf{P}}$=180° is associated with two values of $\Delta\theta_{\mathbf{k}}$: 180° and 60°. When $\Delta\theta_{\mathbf{k}}=180°$, the *W*(180°, 180°) domain walls possess two type of orientations of **n**. For example, if the two domains are $C(111,1\bar{1}0)$ and $C(\bar{1}\bar{1}\bar{1},\bar{1}10)$, and **n** is along $[1\bar{1}0]$, the two **k** vectors are symmetric with respect to the domain wall plane (Fig. 2a), which is thus labelled as *W*(180°, 180°) symmetric walls. On the other hand, when **n** is along $[10\bar{1}]$ or $[01\bar{1}]$, the two **k** vectors are asymmetric relative to the domain wall as shown in Fig. 2(b), labelled as a *W*(180°, 180°) asymmetric wall. Similarly, *W*(180°, 60°) walls can be symmetric or asymmetric, as demonstrated in Figs. 2(c) and 2(d). From Table II, when $\Delta\theta_{\mathbf{P}}$=0°, $\Delta\theta_{\mathbf{k}}$ can only take the value of 120°. The **k** vectors across the $W(0°, 120°)$ walls can be symmetric or asymmetric with respect to the domain walls, as shown in Figs. 2(e) and 2(f).

TABLE II. 14 types of AFM walls in BFO including the magnetic contribution to the domain wall energies. Only the energy terms related to **L** are included for the wall energy calculation. The angle change of the **m** vector across the wall is also labelled for each domain wall.

| **Domain wall types** $W(\Delta\theta_P, \Delta\theta_k)$ | **Angles between m vectors** $\Delta\theta_m$ | **Partial wall energy** ($\mu J/m^2$) |
|---|---|---|
| (0°, 120°) symmetric | 120° | 28.9 |
| (0°, 120°) asymmetric | 120° | 88.3 |
| (180°, 180°) symmetric | 0° | 0.632 |
| (180°, 180°) asymmetric | 0° | 232 |
| (180°, 60°) symmetric | 120° | 173 |
| (180°, 60°) asymmetric | 120° | 99.7 |
| (71°, 0°) | 71° | 72.4 |
| (71°, 60°) | 34° | 25.9 |
| (71°, 120°) | 100° | 328 |
| (71°, 90°) | 132° | 84.2 |
| (109°, 180°) | 71° | 155 |
| (109°, 60°) | 100° | 451 |
| (109°, 120°) | 34° | 93.3 |
| (109°, 90°) | 132° | 90.1 |

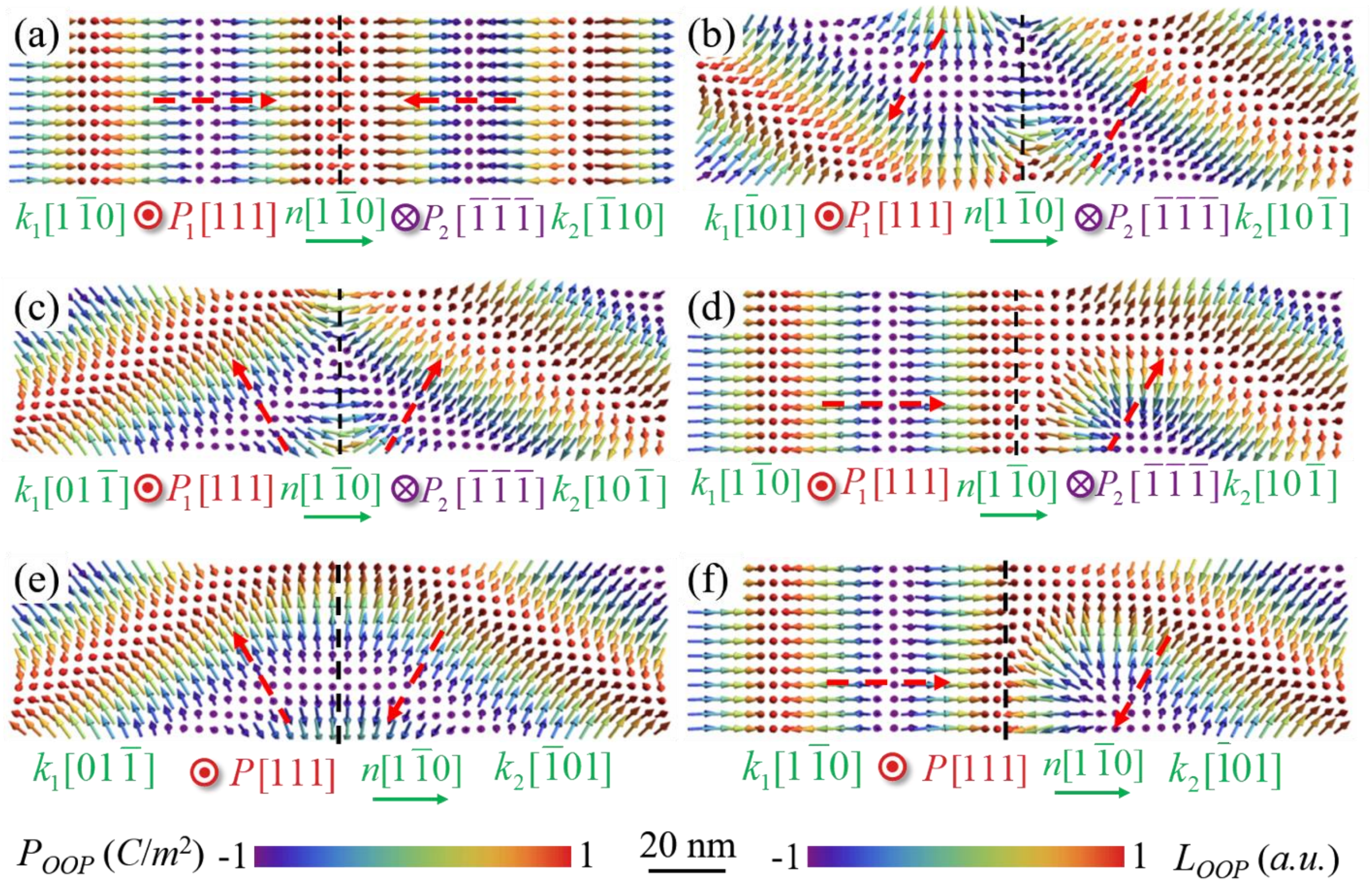

FIG. 2. AFM domain walls with $\Delta\theta_{\mathbf{P}}$ =180° and $\Delta\theta_{\mathbf{P}}$ =0°. Distribution of **L** across a *W*(180°, 180°) symmetric (a), *W*(180°, 180°) asymmetric (b), *W*(180°, 60°) symmetric (c), *W*(180°, 60°) asymmetric wall (d), *W*(0°, 120°) symmetric (e), and *W*(0°, 120°) asymmetric wall (f). The colors of the arrows indicate the components along out of the plane (OOP) of the page. On the bottom of each panel, we label the directions of **P** and **n**. The red dashed arrows specify the **k** direction.

The domain wall orientations with $\Delta\theta_{\mathbf{P}}$ =71° and $\Delta\theta_{\mathbf{P}}$ =109° are constrained by the mechanical compatibility condition and electrical neutrality condition [33], and thus the corresponding low-energy domain walls are uniquely specified by ( $\Delta\theta_{\mathbf{P}}$ , $\Delta\theta_{\mathbf{k}}$ ). When $\Delta\theta_{\mathbf{P}}$ =71°, $\Delta\theta_{\mathbf{k}}$ has four options: 0°, 60°, 90°, and 120°, as listed in Table II. The spin structures across the *W*(71°, 0°), *W*(71°, 60°), *W*(71°, 90°), and *W*(71°, 120°) walls are illustrated in Figs. 3(a)-3(d). Similarly, when $\Delta\theta_{\mathbf{P}}$ =109°, $\Delta\theta_{\mathbf{k}}$ has four options: 180°, 120°, 90°, and 60°, as listed in Table II. The four domain walls with $\Delta\theta_{\mathbf{P}}$ =109° are shown in Figs. 4(a)-4(d). Therefore, there are totally 14 types of domain walls for the cycloidal AFM order in BFO, as listed in Table II.

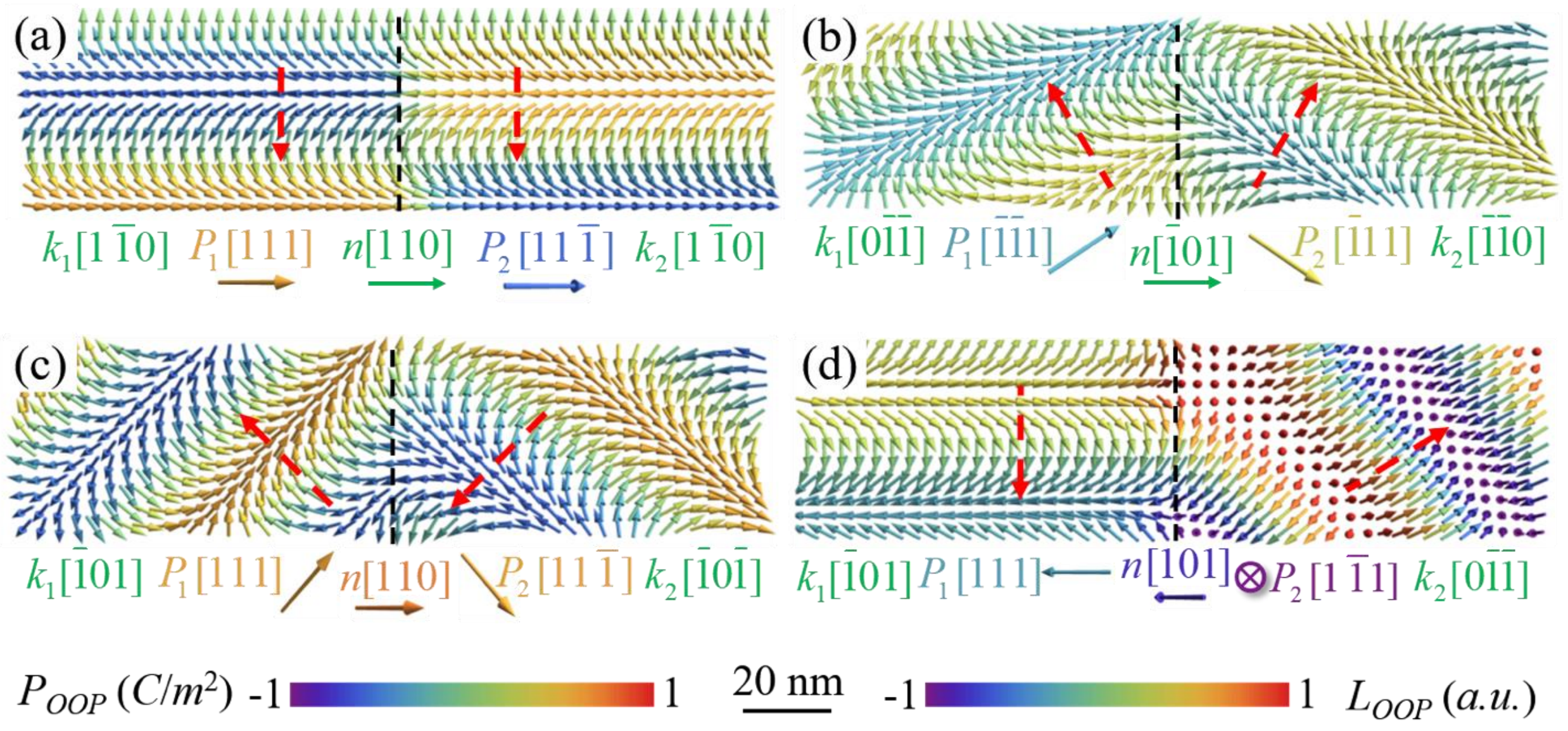

FIG. 3. AFM domain walls with $\Delta\theta_{\mathbf{P}}$ =71°. Distribution of **L** across a *W*(71°, 0°) (a), *W*(71°, 60°) (b), *W*(71°, 90°) (c), and *W*(71°, 120°) wall (d).

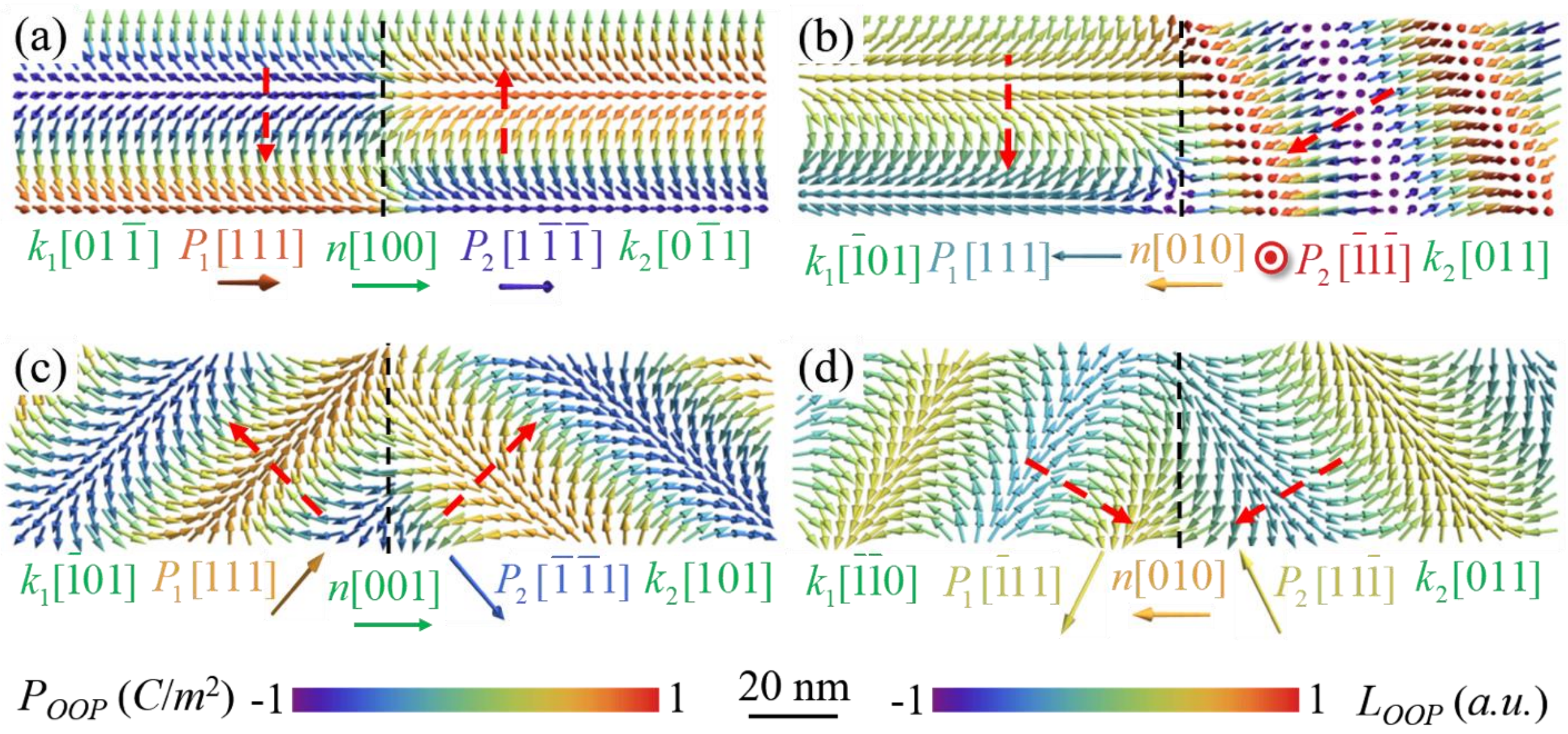

FIG. 4. AFM domain walls with $\Delta\theta_{\mathbf{P}}$ =109°. Distribution of **L** across a *W*(109°, 180°) (a), *W*(109°, 60°) (b), *W*(109°, 90°) (c), and *W*(109°, 120°) wall (d).

## 2.5. Domain wall energies of spin cycloids in BFO

The domain wall spin structures and excess energies are further calculated by the phase-field simulations. Based on the values of $K_1$ and $K_2$ listed in Table SI [23], the difference between the maximum and minimum of the local anisotropy energy density is 2,500 J·m$^{-3}$. The effective uniaxial anisotropy constant is estimated to be $K_u = 2{,}500\ \mathrm{J} \cdot \mathrm{m}^{-3}$, and the magnetocrystalline exchange length is given by $l_{ex} = \sqrt{\dfrac{A}{K_u}} = 40\ \mathrm{nm}$ [34]. Thus it is physically reasonable to choose the grid spacing as $\Delta x = 1.0\ \mathrm{nm}$. Note that $\Delta x$ cannot be set to be too small since the period of the spin cycloid is ~62 nm. In the domain wall simulation, the $BiFeO_3$ region is described by a total area of $600\Delta x \times 600\Delta x \times 1\Delta x$, and buffer layers with a thickness of $202\Delta x$ are added to the boundaries of $BiFeO_3$ to eliminate the effect of periodic boundary conditions (see Fig. S3 [23]). Within the buffer layer, **L** is maintained as **0**. We run phase-field simulations with preset domains of **P**, and the distributions of **L** for the 14 types of domain walls are shown in Figs. 2-4, which are the results after the system is fully relaxed. We compute the system energies from the magnetic contribution for three cases: single domain 1, single domain 2, and two domains with a wall in the middle, whose system energies are labelled by $F_1$, $F_2$, and $F_{12}$, respectively. Then the partial domain wall energy is obtained by $\Delta F = F_{12} - (F_1 + F_2)/2$. Note that only the energy terms dependent on **L** are included in the energy calculation, since the energy penalty from the magnetic part is much smaller than that from the structural part as discussed below.

The obtained partial domain wall energies for the 14 types of domain walls are listed in Table II. The excess energies from the magnetic contribution are on the order of 1~500 $\mu$J/m$^2$, much smaller than the energies from the structural part (50-300 mJ/m$^2$) [19]. Therefore, the total domain wall energy in BFO is mainly determined by the structural part, and the total domain wall energy sequence is $W$(71°, $\Delta\theta_{\mathbf{k}}$)>$W$(180°, $\Delta\theta_{\mathbf{k}}$)>$W$(109°, $\Delta\theta_{\mathbf{k}}$)>$W$(0°, $\Delta\theta_{\mathbf{k}}$) [19,35]. The magnetic part only makes small additional contributions to the total domain wall energies. In Table II, the partial domain wall energy calculation only includes the energy penalty from the magnetic part. The comparison of the partial domain wall energies is meaningful only among the walls with the same $\Delta\theta_{\mathbf{P}}$ and different $\Delta\theta_{\mathbf{k}}$.

As listed in Table II, the energy sequence for $\Delta\theta_{\mathbf{P}}$=71° is $W$(71°, 120°)> $W$(71°, 90°)> $W$(71°, 0°)> $W$(71°, 60°) walls. However, when the thickness of BFO films is on the order of cycloid period, the **k** vectors are constrained within the plane of the film. Thus it is reasonable to observe the *W*(71°, 90°) walls experimentally in (001) BFO thin films [10]. Similarly, in the (001) films with walls $\Delta\theta_{\mathbf{P}}$=109° [36], *W*(109°, 90°) walls are expected. Note that the *W*(180°, 180°) symmetric wall shows a very small excess energy from the magnetic contribution. As shown in Fig. 2(a), the two cycloids possess opposite helicities, with the *W*(180°, 180°) symmetric wall serving as a mirror, resulting in small energy penalty.

2.6. Spin switching dynamics after polarization is switched

The above domain wall structures suggest how the **L** vectors redistribute when **P** rotates. Next we analyze how **L** evolves following the switching of **P** by an electric field. We start with a domain assigned with pre-equilibrium distribution of **P** and **L**, and then we manually switch **P** to $\mathbf{P}'$ at step 0 and temporally and spatially evolve **L**. Similar to the notations of domain walls, a switching path is label by ($\Delta\theta_P$, $\Delta\theta_k$).

Based on whether the initial **k** vector is perpendicular to the switched polarization vector $\mathbf{P}'$ or not, there exist two types of cycloid switchings. When the initial **k** vector is perpendicular to both **P** and $\mathbf{P}'$, the **k** vector either maintains its direction or is switched by 180°. For example, for the domain with $\mathbf{P}=[111]$ and $\mathbf{k}=[1\bar{1}0]$, if $\mathbf{P}'=[11\bar{1}]$, the switched polarization direction $\mathbf{P}'$ is still perpendicular to the initial **k** vector $[1\bar{1}0]$, resulting in a (71°, 0°) switching. Similarly, we have (109°, 180°) and (180°, 180°) switchings. Based on the properties of a spin cycloid as discussed in Section 2.1, a spin cycloid can be assigned with antiparallel **k** vectors, and thus (109°, 180°) and (180°, 180°) switchings can also be described as (109°, 0°) and (180°, 0°) switchings. They are labelled as (109°, 180°) and (180°, 180°) switchings in this paper due to the three rules in Section 2.2. In fact, the **k** vector does not need to be switched by 180° in this case. Note that when $\Delta\theta_{\mathbf{P}}$=180°, the initial **k** is always perpendicular to $\mathbf{P}'$, and only the (180°, 180°) switching is available.

When $\Delta\theta_{\mathbf{P}}$=71° and $\Delta\theta_{\mathbf{P}}$=109°, it is possible that the initial **k** is not perpendicular to $\mathbf{P}'$. Then the **k** vector will be rotated by 60°, 90°, or 120° to reach the low-energy state. From the domain wall types in Table II, 6 types of switchings with $\Delta\theta_{\mathbf{k}} \neq 0$ and $\Delta\theta_{\mathbf{k}} \neq 180°$ are possible,

i.e., (71°, 60°), (71°, 90°), (71°, 120°), (109°, 60°), (109°, 90°), and (109°, 120°) switchings. Overall, 9 types of electrically controlled switching paths are identified.

First, we analyze the situation when the initial **k** vector is perpendicular to both **P** and $\mathbf{P}'$. By solving the LLG equation of a one-dimension (1D) system $1024\Delta x \times 1\Delta x \times 1\Delta x$, we obtain the temporal evolution of **L** for the (71°, 0°) switching as shown in Movie I, and four snapshots are listed in Fig. 5(a). Based on the initial **L** direction in the first snapshot, we label the spin parallel to **P** as $L_P$, and the spin parallel to **k** as $L_k$. As demonstrated in Fig. 5(a), $L_k$ is almost unchanged during the switching, while $L_P$ gradually rotates to the direction of the **P′** vector. In the transient states, i.e., in the second and third snapshots, the distribution of **L** basically maintains the sinusoidal variation. Using the spherical coordinates with the polar axis aligned along the **k** direction and the azimuthal axis along the **P** direction, the change in $L_P$ is plotted in Fig. 5(b). From Fig. 5(b), the switching time of the (71°, 0°) switching is ~20 ps.

The evolution of **L** for the (180°, 180°) switching is demonstrated in Movie II and in Figs. 5(c) and 5(d). Similar to the (71°, 0°) switching, $L_P$ rotates from the direction of the initial **P** vector and towards the direction of **P′**. Note that $L_k$ maintains its direction, although the chosen **k** vector is flipped to its opposite direction. The results for the (109°, 180°) switching are shown in Movie III, and Fig. S4. The (71°, 0°) and (180°, 180°) switchings are experimentally observed in BFO crystals and (111) BFO films, respectively, which are characterized by analyzing the wave vectors of **L** before and after polarization switching [11,37].

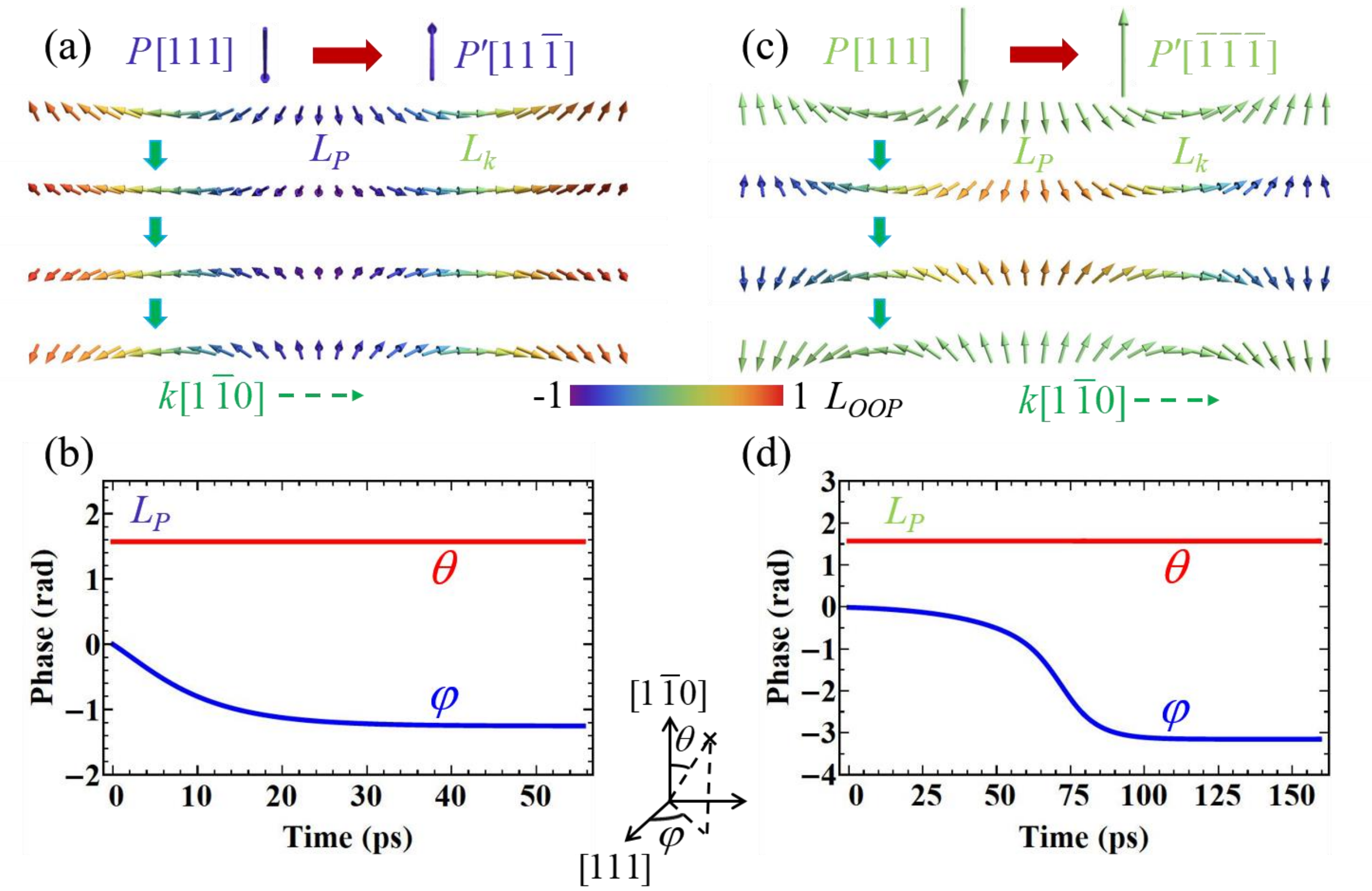

FIG. 5. Spin switching dynamics with $\Delta\theta_{\mathbf{k}} = 0$ or $\Delta\theta_{\mathbf{k}} = 180°$. (a) Four snapshots during the (71°, 0°) switching. $L_P$ is the spin with the initial **L** parallel to **P**. (b) Evolution of $L_P$ during the (71°, 0°) switching based on spherical coordinates. (c) Four snapshots during the (180°, 180°) switching. (d) Evolution of $L_P$ during the (180°, 180°) switching.

To study the spin switching dynamics with the initial **k** vector not perpendicular to $\mathbf{P}'$, 2D simulations are employed. We use the same system setting as in the domain wall energy calculation with buffer layers added at the four boundaries (Fig. S3 [23]). The (71°, 60°) switching is analyzed first, which is found to be characterized by two steps. In step-I, **k** remains unchanged, and **L** rotates locally similar to the (71°, 0°) switching, as demonstrated in Figs. 6(b) and 6(c). However, at the end of step-I, $L_P$ is not aligned along the **P′** direction. Instead, $L_P$ is parallel to the projection of **P′** on the plane perpendicular to the initial **k** vector. To demonstrate this more clearly, we run a 1D simulation in which the **k** vector cannot rotate, and the initial and final spin configurations are shown in Fig. 6(a). Step-II mainly involves the rotation of the **k** vector. As shown in Figs. 6(d) and 6(e), step-II is “first-order” like, and the region with a new **k** vector nucleates and grows at the expense of the region with the initial **k** vector. In the 2D simulation, the boundaries between BFO and buffer layers serve as the nucleation sites as illustrated by Movie IV. The domain wall motion in step-II is slow compared to the local spin flips in step-I. For example, for the (71°, 60°) switching, step-I is completed before 119 ps while step-II takes more than 9,834 ps.

The simulation results for the (71°, 90°), (109°, 120°), and (109°, 90°) switchings are shown in Movies V-VII and Figs. S5-S7. All the three processes exhibit two-step dynamics similar to that of the (71°, 60°) switching. The (71°, 60°), (71°, 90°), (109°, 120°), and (109°, 90°) switchings have one-to-one correspondence with the domain walls in Figs. 3(b), 3(c), 4(d), and 4(c). Similarly, the domain walls in Figs. 3(d) and 4(b) correspond to (71°, 120°) and (109°, 60°) switchings, which are not presented here for simplicity. Therefore, when the **P** vector is switched by 71° and 109°, with the initial **k** vector not perpendicular to **P′**, several switching paths for the **k** vector are possible.

When the **k** vector is constrained within a 2D plane, i.e., in a BFO thin film, with the initial **k** vector not perpendicular to **P′**, the choice of the switching path is dependent on the normal of the 2D plane. For example, with $\Delta\theta_{\mathbf{P}} = 71°$ and with the initial **k** not perpendicular to **P′**, (71°, 60°), (71°, 90°), and (71°, 120°) switchings are all possible. For the situation with $\mathbf{P} = [111]$, $\mathbf{k} = [1\bar{1}0]$, and $\mathbf{P}' = [\bar{1}11]$, if the film normal is along [001], then $\mathbf{k}' = [\bar{1}\bar{1}0]$, resulting in a (71°, 90°) switching; if the film normal is along $[11\bar{1}]$, then $\mathbf{k}' = [101]$, resulting in a (71°, 60°) switching; if the film normal is along $[111]$, then $\mathbf{k}' = [01\bar{1}]$, resulting in a (71°, 120°) switching. Similarly, when $\Delta\theta_{\mathbf{P}} = 109°$, with the initial **k** vector not perpendicular to **P′**, (109°, 60°), (109°, 90°), and (109°, 120°) switchings are expected to occur based on the normal of the 2D plane. This is consistent with experiments, where the (109°, 90°) switching is observed in (001) BFO films [10].

When the **k** vector is free to rotate in the 3D space, i.e., in BFO bulks, with the initial **k** vector not perpendicular to **P′**, phase-field simulations show that the switching is along a path with the smallest **m** vector rotation angle when several options are permissible. We run a 3D simulation with grids $256\Delta x \times 256\Delta x \times 256\Delta x$, which shows that the (71°, 60°) switching is kinetically favored over the (71°, 90°) and (71°, 120°) switchings. Also, with $\Delta\theta_{\mathbf{P}} = 109°$ and with the initial **k** not perpendicular to **P′**, the (109°, 120°) switching is observed in 3D simulations. As demonstrated in Table II, the (71°, 60°) and (109°, 120°) switchings exhibit the smallest rotation angle for the **m** vector with fixed $\Delta\theta_{\mathbf{P}}$.

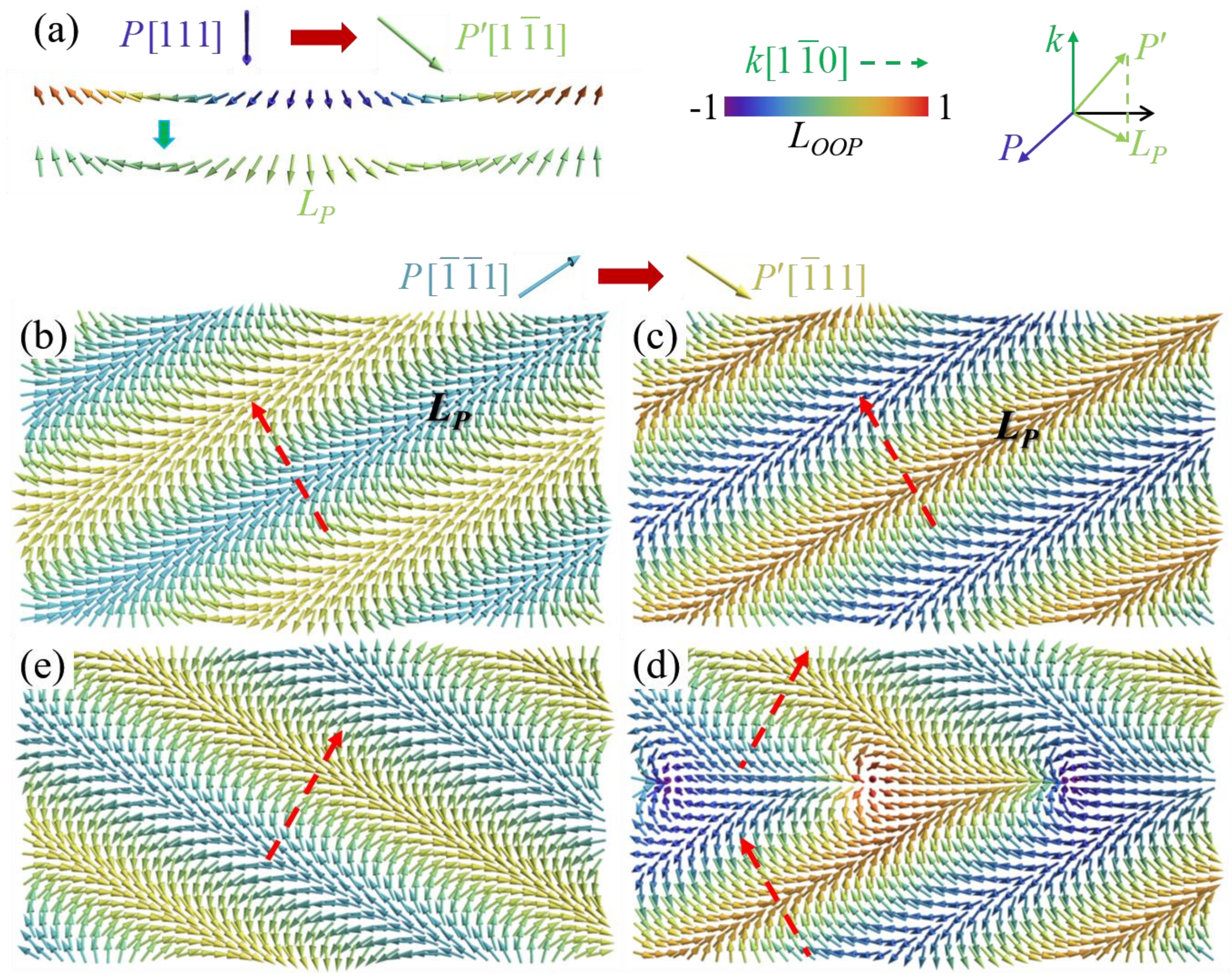

FIG. 6. Spin evolution during the (71°, 60°) switching. (a) Evolution of **L** based on a 1D simulation. (b)-(e) Zoomed snapshots from a 2D simulation, which are the results at (b) 0 ps, (c) 119 ps, (d) 796 ps, and (e) 9,834 ps. The red dashed arrows specify the **k** vector.

### 2.7 Multi-step applications of electric fields

As discussed in Section 2.6, when the **k** vector is free to rotate in the 3D space, only (71°, 0°), (71°, 60°), (109°, 0°), (109°, 120°), and (180°, 180°) switchings are possible. Thus, through a one-step electric field application to a spin cycloid, the number of the resulting spin cycloids is limited. However, we will show that by rotating the **P** vector for multiple times consecutively, the switching between any two of the 24 cycloids is achievable.

In BFO films, it is shown that the 180° polarization switching itself is a two-step process, i.e., a 71° switching followed by a 109° switching, due to the high energy barrier associated with the direct 180° polarization switching [38,39]. In BFO single crystals, only 71° polarization switching is observed in experiments [11]. Therefore, in the following discussion, we first assume that only 71° polarization switching is available, and then assume that both 71° and 109° polarization switchings are possible.

When only 71° polarization switching is available, within four steps of electric field application, it is possible to switch one spin cycloid to any of the 23 other cycloids, as demonstrated in Fig. 7. For the first step switching, there are three possibilities since one of the three polarization components can be switched. For the second and third step switchings, only two options are possible to each cycloid since the third one will switch the spin cycloid back to the cycloid before the previous switching. For the forth step switching, only one possible path for each resulting cycloid is labeled for clarity.

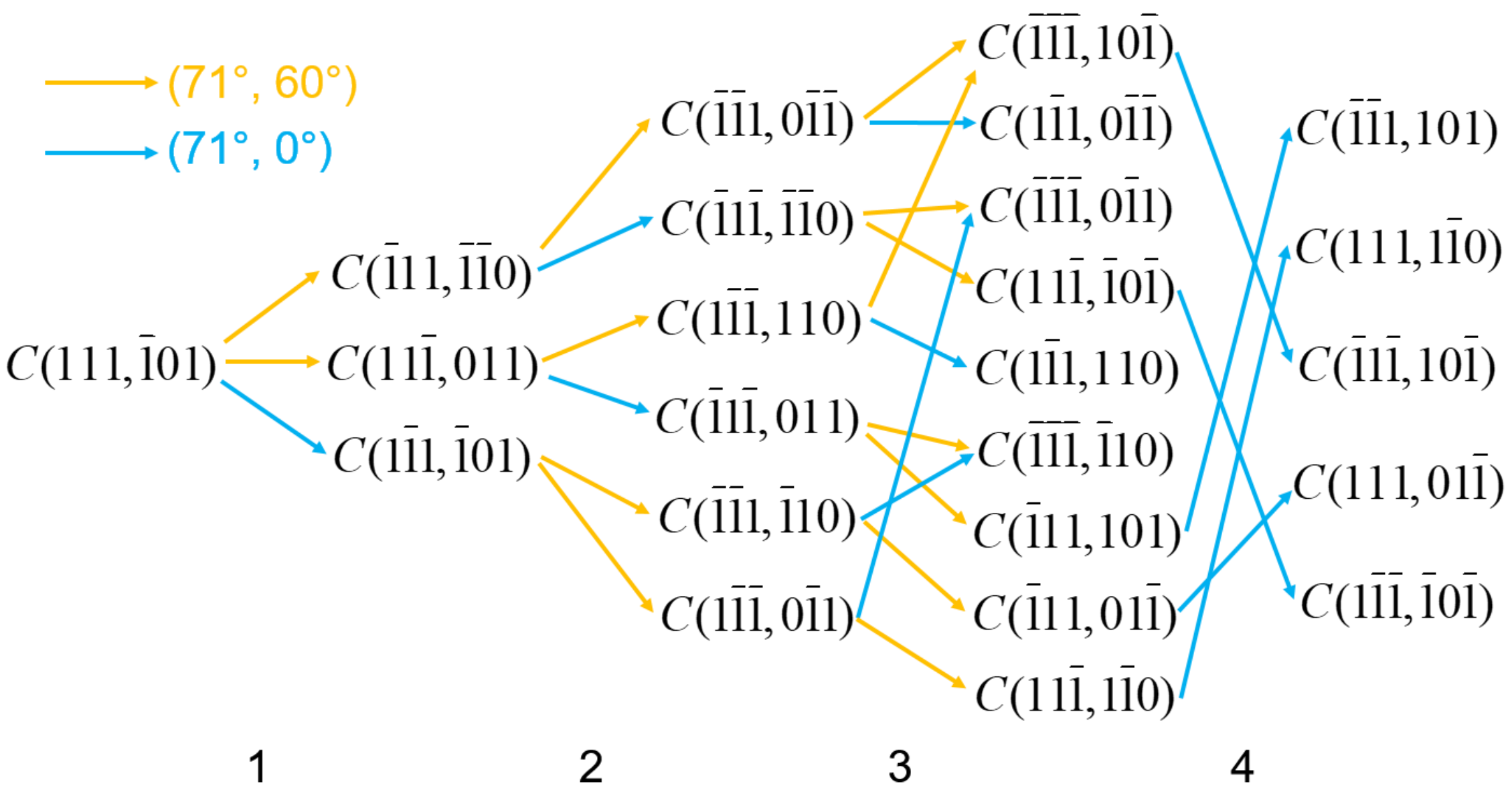

FIG. 7. Roadmap for spin cycloid switching if only 71° polarization switching is available. The arrows with different colors indicate the switching paths for spin cycloids. A spin cycloid is labeled by C(**P**, **k**).

When both 71° and 109° polarization switchings are available, within three steps of electric field application, it is possible to switch one spin cycloid to any of the 23 other cycloids, as illustrated in Fig. 8. For the first step switching, six options are possible. For the second and third step switchings, only one possible path for each resulting cycloid is labeled for clarity. Note that in both Figs. 7 and 8, the number of switching steps is minimized for a spin cycloid, i.e., only the simplest path is plotted.

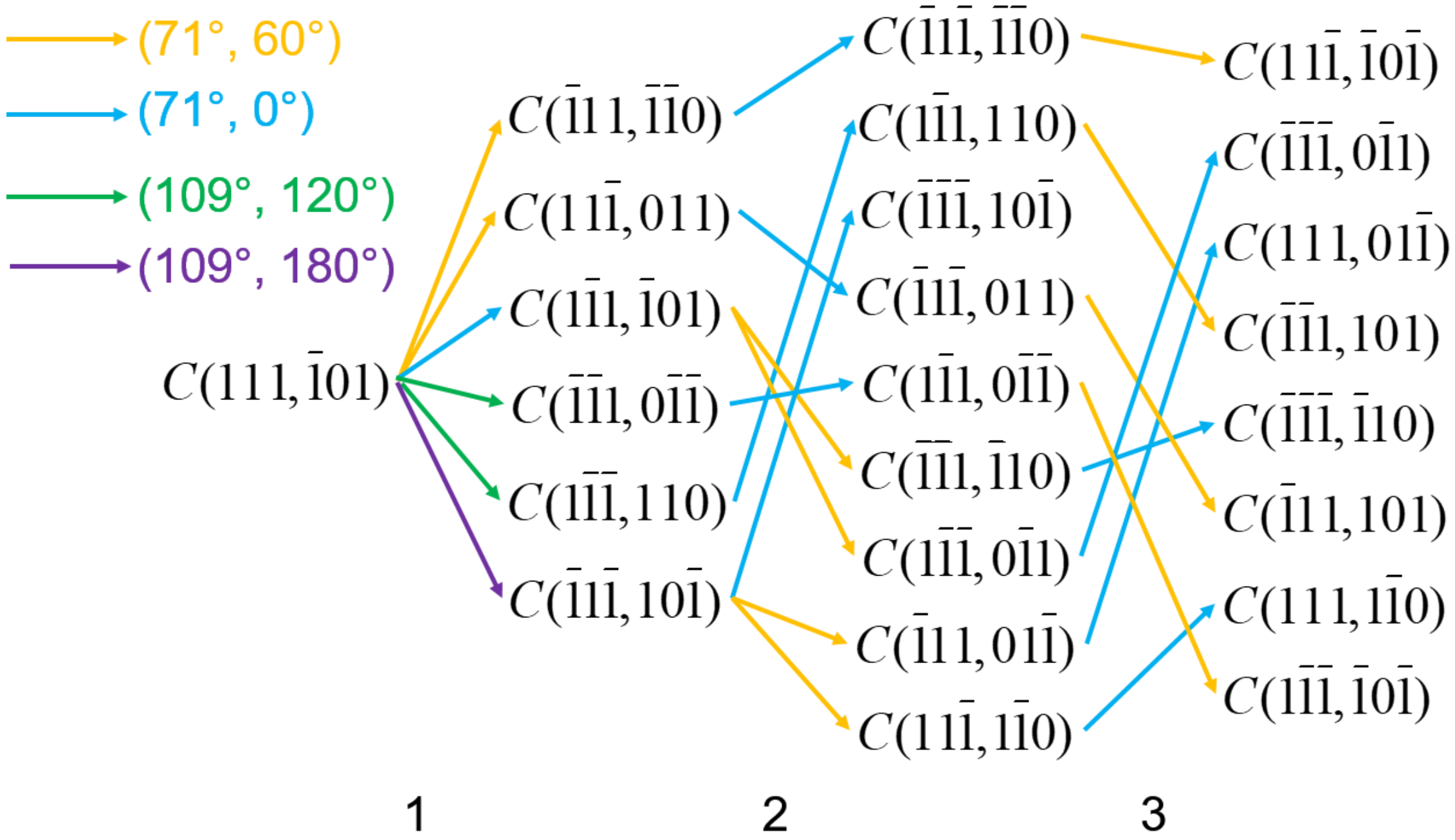

FIG. 8. Roadmap for spin cycloid switching if both 71° and 109° polarization switchings are available. The arrows with different colors indicate the switching paths for spin cycloids.

## 3. Conclusions

We employed Landau theory and phase-field simulations to systematically analyze the magnetic domain walls and electrically controlled spin switching dynamics in the cycloidal antiferromagnetic state of $BiFeO_3$. We identified 14 types of magnetic domain walls and obtained the corresponding partial domain wall energies from the magnetic contribution. We discovered 9 paths of electric-field-induced spin switching, where the **k** vector direction is rotated by 0°, 60°, 90°, 120°, and 180°. Interestingly, when the **k** vector is rotated by 60°, 90°, and 120°, the switching process involves two distinct steps: fast local flips followed by slow rotation of wave vectors. We also demonstrated that through multi-step applications of electric fields, it is possible to achieve the switching between any two spin cycloids. Since cycloidal spin orders are common in multiferroics while $BiFeO_3$ is the best known multiferroics being explored in multiferroic/ferromagnet heterostructures [40,41], our work provides a new theoretical framework to understand and design the mutual control of polarization and magnetic orders.

**Acknowledgement:**

The work is supported as part of the Computational Materials Sciences Program funded by the U.S. Department of Energy, Office of Science, Basic Energy Sciences, under Award No. DE-SC0020145. Computations for this research were partially performed on the Pennsylvania State University's Institute for Computational and Data Sciences' Roar supercomputer. This work used the Extreme Science and Engineering Discovery Environment (XSEDE), which is supported by National Science Foundation grant number ACI-1548562. The work used XSEDE Comet at San Diego Supercomputer Center through allocation TG-DMR170006.

**References:**

[1] S.-W. Cheong and M. Mostovoy, *Multiferroics: a magnetic twist for ferroelectricity*, Nature materials **6**, 13 (2007).
[2] R. Ramesh and N. A. Spaldin, *Multiferroics: progress and prospects in thin films*, Nature materials **6**, 21 (2007).
[3] M. Mostovoy, *Ferroelectricity in spiral magnets*, Physical Review Letters **96**, 067601 (2006).
[4] Y. Tokura and S. Seki, *Multiferroics with spiral spin orders*, Advanced materials **22**, 1554 (2010).
[5] Y. Yamasaki, S. Miyasaka, Y. Kaneko, J.-P. He, T. Arima, and Y. Tokura, *Magnetic reversal of the ferroelectric polarization in a multiferroic spinel oxide*, Physical review letters **96**, 207204 (2006).
[6] Y. Yamasaki, H. Sagayama, T. Goto, M. Matsuura, K. Hirota, T. Arima, and Y. Tokura, *Electric control of spin helicity in a magnetic ferroelectric*, Physical review letters **98**, 147204 (2007).
[7] G. Catalan and J. F. Scott, *Physics and applications of bismuth ferrite*, Advanced Materials **21**, 2463 (2009).
[8] D. Lebeugle, D. Colson, A. Forget, M. Viret, P. Bonville, J.-F. Marucco, and S. Fusil, *Room-temperature coexistence of large electric polarization and magnetic order in Bi Fe O 3 single crystals*, Physical Review B **76**, 024116 (2007).
[9] S. Dong, J.-M. Liu, S.-W. Cheong, and Z. Ren, *Multiferroic materials and magnetoelectric physics: symmetry, entanglement, excitation, and topology*, Advances in Physics **64**, 519 (2015).

[10] I. Gross, W. Akhtar, V. Garcia, L. Martínez, S. Chouaieb, K. Garcia, C. Carrétéro, A. Barthélémy, P. Appel, and P. Maletinsky, *Real-space imaging of non-collinear antiferromagnetic order with a single-spin magnetometer*, Nature **549**, 252 (2017).
[11] D. Lebeugle, D. Colson, A. Forget, M. Viret, A. M. Bataille, and A. Gukasov, *Electric-field-induced spin flop in BiFeO 3 single crystals at room temperature*, Physical review letters **100**, 227602 (2008).
[12] I. Sosnowska, T. P. Neumaier, and E. Steichele, *Spiral magnetic ordering in bismuth ferrite*, Journal of Physics C: Solid State Physics **15**, 4835 (1982).
[13] H. Béa, M. Bibes, S. Petit, J. Kreisel, and A. Barthélémy, *Structural distortion and magnetism of BiFeO3 epitaxial thin films: a Raman spectroscopy and neutron diffraction study*, Philosophical Magazine Letters **87**, 165 (2007).
[14] J. Heron, J. Bosse, Q. He, Y. Gao, M. Trassin, L. Ye, J. Clarkson, C. Wang, J. Liu, and S. Salahuddin, *Deterministic switching of ferromagnetism at room temperature using an electric field*, Nature **516**, 370 (2014).
[15] D. Sando, A. Agbelele, D. Rahmedov, J. Liu, P. Rovillain, C. Toulouse, I. Infante, A. Pyatakov, S. Fusil, and E. Jacquet, *Crafting the magnonic and spintronic response of BiFeO 3 films by epitaxial strain*, Nature materials **12**, 641 (2013).
[16] W. Saenrang, B. Davidson, F. Maccherozzi, J. Podkaminer, J. Irwin, R. Johnson, J. Freeland, J. Íñiguez, J. Schad, and K. Reierson, *Deterministic and robust room-temperature exchange coupling in monodomain multiferroic BiFeO 3 heterostructures*, Nature Communications **8**, 1583 (2017).
[17] J. Bertinshaw, R. Maran, S. J. Callori, V. Ramesh, J. Cheung, S. A. Danilkin, W. T. Lee, S. Hu, J. Seidel, and N. Valanoor, *Direct evidence for the spin cycloid in strained nanoscale bismuth ferrite thin films*, Nature communications **7**, 12664 (2016).
[18] J.-C. Toledano and P. Toledano, *The Landau theory of phase transitions: application to structural, incommensurate, magnetic and liquid crystal systems* (World Scientific Publishing Company, 1987), Vol. 3.
[19] F. Xue, Y. Gu, L. Liang, Y. Wang, and L.-Q. Chen, *Orientations of low-energy domain walls in perovskites with oxygen octahedral tilts*, Physical Review B **90**, 220101(R) (2014).
[20] R. de Sousa and J. E. Moore, *Optical coupling to spin waves in the cycloidal multiferroic Bi Fe O 3*, Physical Review B **77**, 012406 (2008).
[21] F. Xue, L. Li, J. Britson, Z. Hong, C. A. Heikes, C. Adamo, D. G. Schlom, X. Pan, and L.-Q. Chen, *Switching the curl of polarization vectors by an irrotational electric field*, Physical Review B **94**, 100103(R) (2016).
[22] A. Hubert and R. Schäfer, *Magnetic domains: the analysis of magnetic microstructures* (Springer Science & Business Media, 2008).
[23] See supplemenary materials for captions of movies, full expression of Eq. (1), free energy coefficients, derivation of Eq. (3), derivation of Eq. (4), remark on cycloid wavelength, remark on modulation directions, 2D system setting, and spin evolution during the (109°, 180°), (71°, 90°), (109°, 120°), and (109°, 90°) switchings.
[24] L. Q. Chen and J. Shen, *Applications of semi-implicit Fourier-spectral method to phase field equations*, Computer Physics Communications **108**, 147 (1998).
[25] F. Xue, X. Wang, Y. Shi, S.-W. Cheong, and L.-Q. Chen, *Strain-induced incommensurate phases in hexagonal manganites*, Physical Review B **96**, 104109 (2017).

[26] J. Towns, T. Cockerill, M. Dahan, I. Foster, K. Gaither, A. Grimshaw, V. Hazlewood, S. Lathrop, D. Lifka, and G. D. Peterson, *XSEDE: accelerating scientific discovery*, Computing in Science & Engineering **16**, 62 (2014).
[27] F. Xue, N. Wang, X. Wang, Y. Ji, S.-W. Cheong, and L.-Q. Chen, *Topological dynamics of vortex-line networks in hexagonal manganites*, Physical Review B **97**, 020101(R) (2018).
[28] V. Baltz, A. Manchon, M. Tsoi, T. Moriyama, T. Ono, and Y. Tserkovnyak, *Antiferromagnetic spintronics*, Reviews of Modern Physics **90**, 015005 (2018).
[29] K. M. Hals, Y. Tserkovnyak, and A. Brataas, *Phenomenology of current-induced dynamics in antiferromagnets*, Physical review letters **106**, 107206 (2011).
[30] J. Lu, A. Günther, F. Schrettle, F. Mayr, S. Krohns, P. Lunkenheimer, A. Pimenov, V. Travkin, A. Mukhin, and A. Loidl, *On the room temperature multiferroic BiFeO 3: magnetic, dielectric and thermal properties*, The European Physical Journal B **75**, 451 (2010).
[31] I. Sosnowska, W. Schäfer, W. Kockelmann, K. Andersen, and I. Troyanchuk, *Crystal structure and spiral magnetic ordering of BiFeO 3 doped with manganese*, Applied Physics A **74**, s1040 (2002).
[32] J. Zhang and L. Chen, *Phase-field microelasticity theory and micromagnetic simulations of domain structures in giant magnetostrictive materials*, Acta Materialia **53**, 2845 (2005).
[33] P. Marton, I. Rychetsky, and J. Hlinka, *Domain walls of ferroelectric BaTiO 3 within the Ginzburg-Landau-Devonshire phenomenological model*, Physical Review B **81**, 144125 (2010).
[34] G. S. Abo, Y.-K. Hong, J. Park, J. Lee, W. Lee, and B.-C. Choi, *Definition of magnetic exchange length*, IEEE Transactions on Magnetics **49**, 4937 (2013).
[35] Y. Wang, C. Nelson, A. Melville, B. Winchester, S. Shang, Z.-K. Liu, D. G. Schlom, X. Pan, and L.-Q. Chen, *BiFeO 3 domain wall energies and structures: a combined experimental and density functional theory+ U study*, Physical review letters **110**, 267601 (2013).
[36] Y.-H. Chu, Q. He, C.-H. Yang, P. Yu, L. W. Martin, P. Shafer, and R. Ramesh, *Nanoscale control of domain architectures in BiFeO3 thin films*, Nano letters **9**, 1726 (2009).
[37] N. W. Price, A. Vibhakar, R. Johnson, J. Schad, W. Saenrang, A. Bombardi, F. Chmiel, C. Eom, and P. Radaelli, *Strain engineering a multiferroic monodomain in thin-film BiFeO $ _3$*, arXiv preprint arXiv:1808.10666 (2018).
[38] S. Baek, H. Jang, C. Folkman, Y. Li, B. Winchester, J. Zhang, Q. He, Y. Chu, C. Nelson, and M. Rzchowski, *Ferroelastic switching for nanoscale non-volatile magnetoelectric devices*, Nature materials **9**, 309 (2010).
[39] Y.-H. Hsieh, F. Xue, T. Yang, H.-J. Liu, Y. Zhu, Y.-C. Chen, Q. Zhan, C.-G. Duan, L.-Q. Chen, and Q. He, *Permanent ferroelectric retention of BiFeO 3 mesocrystal*, Nature communications **7**, 13199 (2016).
[40] P. Yu, Y. Chu, and R. Ramesh, *Emergent phenomena at multiferroic heterointerfaces*, Phil. Trans. R. Soc. A **370**, 4856 (2012).
[41] J. Heron, D. Schlom, and R. Ramesh, *Electric field control of magnetism using BiFeO3-based heterostructures*, Applied Physics Reviews **1**, 021303 (2014).

Supplementary Materials: Theory and phase-field simulations on electrical control of spin cycloids in a multiferroic

Fei Xue,* Tiannan Yang, and Long-Qing Chen

Department of Materials Science and Engineering, The Pennsylvania State University, University Park, Pennsylvania 16802, USA

*Corresponding author: xuefei5376@gmail.com

**Supplementary movies**

Supplementary Movie I: Evolution of the antiferromagnetic order parameter **L** during the (71°, 0°) switching. The dimensionless time step $\Delta\tau = 0.01$ (real time step $\Delta t$ is related to $\Delta\tau$ through $\Delta t = \frac{\nu\mu_0 L_s}{\omega f_0}\Delta\tau$ , where $f_0$ is the normalization constant for energy density given by $f_0 = 1.0\times10^6\ J\cdot m^{-3}$). We show one frame after each 500 time steps.

Supplementary Movie II: Evolution of the antiferromagnetic order parameter **L** during the (180°, 180°) switching. The dimensionless time step $\Delta\tau = 0.01$. We show one frame after each 500 time steps.

Supplementary Movie III: Evolution of the antiferromagnetic order parameter **L** during the (109°, 180°) switching. The dimensionless time step $\Delta\tau = 0.01$. We show one frame after each time 500 steps.

Supplementary Movie IV: Evolution of the antiferromagnetic order parameter **L** during the (71°, 60°) switching. The switching process is slow, and the dimensionless time step is set to $\Delta\tau = 0.05$. To reduce the file size, we choose different simulation steps between neighboring frames since the evolution becomes slower with larger simulation steps. 0-5, 000 steps, we choose one frame after each 500 steps; 5,000-20,000 steps, we choose one frame after each 3,000 steps; 20,000-120,000 steps, we choose one frame after each 20,000 steps.

Supplementary Movie V: Evolution of the antiferromagnetic order parameter **L** during the (71°, 90°) switching. The process is slow, and the dimensionless time step is set to $\Delta\tau = 0.05$. To reduce the file size, we choose different simulation steps between neighboring frames. 0-5, 000 steps, we choose one frame after each 500 steps; 5,000-20,000 steps, we choose one frame after each 3,000 steps; 20,000-120,000 steps, we choose one frame after each 20,000 steps.

Supplementary Movie VI: Evolution of the antiferromagnetic order parameter **L** during the (109°, 120°) switching. The process is slow, and the dimensionless time step is set to $\Delta\tau = 0.05$. To reduce the file size, we choose different simulation steps between neighboring frames. 0-5, 000 steps, we choose one frame after each 500 steps; 5,000-20,000 steps, we choose one frame after each 3,000 steps; 20,000-120,000 steps, we choose one frame after each 20,000 steps.

Supplementary Movie VII: Evolution of the antiferromagnetic order parameter **L** during the (109°, 90°) switching. The process is slow, and the dimensionless time step is set to $\Delta\tau = 0.05$. To reduce the file size, we choose different simulation steps between neighboring frames. 0-5, 000 steps, we choose one frame after each 500 steps; 5,000-20,000 steps, we choose one frame after each 3,000 steps; 20,000-120,000 steps, we choose one frame after each 20,000 steps.

## I. Full expression of Eq. (1) in the main text

The full expression of Eq. (1) in the main text is given by:

$$\begin{aligned}
F = \int dV \Big\langle & \alpha_{11}(P_1^2 + P_2^2 + P_3^2) + \alpha_{1111}(P_1^4 + P_2^4 + P_3^4) + \alpha_{1122}[P_2^2P_3^2 + P_1^2(P_2^2 + P_3^2)] \\
& + \beta_{11}(\theta_1^2 + \theta_2^2 + \theta_3^2) + \beta_{1111}(\theta_1^4 + \theta_2^4 + \theta_3^4) + \beta_{1122}[\theta_2^2\theta_3^2 + \theta_1^2(\theta_2^2 + \theta_3^2)] \\
& + t_{1111}(P_1^2\theta_1^2 + P_2^2\theta_2^2 + P_3^2\theta_3^2) + t_{1122}[P_1^2(\theta_2^2 + \theta_3^2) + P_2^2(\theta_1^2 + \theta_3^2) + P_3^2(\theta_1^2 + \theta_2^2)] \\
& + t_{1212}(P_1P_2\theta_1\theta_2 + P_2P_3\theta_2\theta_3 + P_3P_1\theta_3\theta_1) + K_1(L_1^2L_2^2 + L_1^2L_3^2 + L_2^2L_3^2) + K_2L_1^2L_2^2L_3^2 \\
& + h(L_1P_1 + L_2P_2 + L_3P_3)^2 + \frac{1}{2}g_{1111}(P_{1,1}^2 + P_{2,2}^2 + P_{3,3}^2) + g_{1122}(P_{1,1}P_{2,2} + P_{2,2}P_{3,3} + P_{1,1}P_{3,3}) \\
& + \frac{1}{2}g_{1212}[(P_{1,2} + P_{2,1})^2 + (P_{2,3} + P_{3,2})^2 + (P_{1,3} + P_{3,1})^2] + \frac{1}{2}\kappa_{1111}(\theta_{1,1}^2 + \theta_{2,2}^2 + \theta_{3,3}^2) \\
& + \kappa_{1122}(\theta_{1,1}\theta_{2,2} + \theta_{2,2}\theta_{3,3} + \theta_{1,1}\theta_{3,3}) + \frac{1}{2}\kappa_{1212}[(\theta_{1,2} + \theta_{2,1})^2 + (\theta_{2,3} + \theta_{3,2})^2 \\
& + (\theta_{1,3} + \theta_{3,1})^2] + A(L_{1,1}^2 + L_{1,2}^2 + L_{1,3}^2 + L_{2,1}^2 + L_{2,2}^2 + L_{2,3}^2 + L_{3,1}^2 + L_{3,2}^2 + L_{3,3}^2) \\
& + \gamma\{L_1[P_1(L_{2,2} + L_{3,3}) - P_2L_{2,1} - P_3L_{3,1}] + L_2[P_2(L_{3,3} + L_{1,1}) - P_1L_{1,2} - P_3L_{3,2}] \\
& + L_3[P_3(L_{1,1} + L_{2,2}) - P_1L_{1,3} - P_2L_{2,3}]\}\Big\rangle
\end{aligned} \quad , \quad \text{(S1)}$$

where a comma in the subscript represents a spatial derivative.

## II. Coefficients of $BiFeO_3$ used in the phase-field simulations

TABLE SI. Coefficients of $BiFeO_3$ used in the simulations (SI units)

| | | | |
|---|---|---|---|
| $\alpha_{11}$ | $-3.580\times10^{8}$ $\mathrm{C^{-2}\cdot m^{2}\cdot N}$ | $\kappa_{1111}$ | $7.840\times10^{-11}$ $\mathrm{rad^{-2}\cdot N}$ |
| $\alpha_{1111}$ | $3.000\times10^{8}$ $\mathrm{C^{-4}\cdot m^{6}\cdot N}$ | $\kappa_{1122}$ | $-5.138\times10^{-9}$ $\mathrm{rad^{-2}\cdot N}$ |
| $\alpha_{1122}$ | $1.188\times10^{8}$ $\mathrm{C^{-4}\cdot m^{6}\cdot N}$ | $\kappa_{1212}$ | $4.977\times10^{-9}$ $\mathrm{rad^{-2}\cdot N}$ |
| $\beta_{11}$ | $-5.400\times10^{9}$ $\mathrm{rad^{-2}\cdot m^{-2}\cdot N}$ | $h$ | $-3.2\times10^{4}$ $\mathrm{C^{-2}\cdot m^{2}\cdot N}$ |
| $\beta_{1111}$ | $3.440\times10^{10}$ $\mathrm{rad^{-4}\cdot m^{-2}\cdot N}$ | $A$ | $4.0\times10^{-12}$ $\mathrm{J\cdot m^{-1}}$ |
| $\beta_{1122}$ | $6.799\times10^{10}$ $\mathrm{rad^{-4}\cdot m^{-2}\cdot N}$ | $\gamma$ | $8.1\times10^{-4}$ $\mathrm{J\cdot C^{-1}}$ |
| $t_{1111}$ | $4.532\times10^{9}$ $\mathrm{C^{-2}\cdot rad^{-2}\cdot m^{2}\cdot N}$ | $K_1$ | $-1.0\times10^{4}$ $\mathrm{J\cdot m^{-3}}$ |
| $t_{1122}$ | $2.266\times10^{9}$ $\mathrm{C^{-2}\cdot rad^{-2}\cdot m^{2}\cdot N}$ | $K_2$ | $3.1\times10^{4}$ $\mathrm{J\cdot m^{-3}}$ |
| $t_{1212}$ | $-4.840\times10^{9}$ $\mathrm{C^{-2}\cdot rad^{-2}\cdot m^{2}\cdot N}$ | $L_S$ | $5.6\times10^{5}$ $\mathrm{A\cdot m^{-1}}$ |
| $g_{1111}$ | $4.335\times10^{-11}$ $\mathrm{C^{-2}\cdot m^{4}\cdot N}$ | $\omega$ | $2.21\times10^{5}$ $\mathrm{m\cdot A^{-1}\cdot s^{-1}}$ |
| $g_{1122}$ | $-3.400\times10^{-12}$ $\mathrm{C^{-2}\cdot m^{4}\cdot N}$ | $\upsilon$ | 0.5 |
| $g_{1212}$ | $3.400\times10^{-12}$ $\mathrm{C^{-2}\cdot m^{4}\cdot N}$ | | |

## III. Derivation of Eq. (3) for the evolution of L

For the derivation purpose, we temporarily introduce the total magnetization order parameter **M**, which is defined by $\mathbf{M}=\mathbf{M}_a+\mathbf{M}_b$ [1,2]. Although **M** is much smaller than **L** and thus neglected in our simulations, it is useful in deriving the evolution equation of **L,** i.e., Eq. (3) of the main text. In the free energy density Eq. (1) of the main text, a new term related to **M** is added, i.e., $a\mathbf{M}^2/2$, where $a$ is the Landau coefficient related to magnetic susceptibility [1]. Following Eq. (8) of Ref. [1], the evolution equation for **L** to the linear order without current injection and external magnetic field is given by

$$\ddot{\mathbf{L}}/\tilde{\omega}=\upsilon\dot{\mathbf{f}}_{\mathrm{L}}+a(\omega\mathbf{f}_{\mathrm{L}}-\upsilon\dot{\mathbf{L}}),\tag{S2}$$

where $\omega$ is the electron gyromagnetic ratio and $\nu$ the Gilbert damping constant. $\tilde{\omega}=\frac{\omega}{1+\upsilon^2}$.

$\mathbf{f}_\mathbf{L}=-\frac{1}{\mu_0 L_S}\frac{\delta F}{\delta \mathbf{L}}$. Next we will demonstrate that since $a$ is large in BFO, the terms $\ddot{\mathbf{L}}/\tilde{\omega}$ and $\upsilon\dot{\mathbf{f}}_\mathbf{L}$ can be neglected if we focus on the evolution at a time scale larger than 1 ps.

Experimental measurements show that the magnetic susceptibility of BFO is $\chi = 7.37\times10^{-4}$ [3]. We obtain $a=\mu_0 L_S^2\frac{1}{\chi}=5.35\times10^8\ \mathrm{N\cdot m^{-2}}$, where $\mu_0$ is the vacuum permeability and $L_S$ is the saturation AFM magnetization ($L_S \sim 5.6\times10^5\ \mathrm{A\cdot m^{-1}}$ following [4]). Using the typical values of $\omega$ and $\nu$ with $\omega = 2.21\times10^5\ \mathrm{m\cdot A^{-1}\cdot s^{-1}}$ and $\upsilon=0.5$ [5], when the characteristic time is $\tilde{t}=1\ \mathrm{ps}$, we have $\frac{\ddot{\mathbf{L}}}{\tilde{\omega}}/a\upsilon\dot{\mathbf{L}} \sim \frac{\dot{\mathbf{L}}}{\tilde{t}\tilde{\omega}}/a\upsilon\dot{\mathbf{L}} \sim 2.11\times10^{-2}$ and $\upsilon\dot{\mathbf{f}}_\mathbf{L}/a\omega\mathbf{f}_\mathbf{L} \sim \frac{\upsilon\mathbf{f}_\mathbf{L}}{\tilde{t}}/a\omega\mathbf{f}_\mathbf{L} \sim 4.23\times10^{-3}$, which indicates that the terms $\ddot{\mathbf{L}}/\tilde{\omega}$ and $\upsilon\dot{\mathbf{f}}_\mathbf{L}$ are relative small and can be ignored in this time scale. Then Eq. (S2) is further simplified to

$$\dot{\mathbf{L}}=\frac{\omega}{\upsilon}\mathbf{f}_\mathbf{L}, \tag{S3}$$

Eq. (S3) has the same form as the TDGL equation. However, different from the typical order parameters described by the TDGL equation which can change both their directions and magnitude, the order parameter **L** in Eq. (S3) maintains its magnitude. Note that Eq. (S3) is the evolution equation at a time scale coarser than 1 ps. At a time scale finer than 0.1 ps, the neglected terms becomes important, and the dynamics is more complex [1,2].

## IV. Derivation of cycloidal distribution Eq. (4)

In a spin cycloid, **L** rotates within a two-dimensional (2D) plane, and we assume the plane is spanned by the unit vectors $\mathbf{e}_1$ and $\mathbf{e}_2$. Since **P** is within the 2D plane, we can choose $\mathbf{e}_1$ along the **P** direction, i.e., $\mathbf{e}_1=\frac{\mathbf{P}}{|\mathbf{P}|}$. Note that in 3D spaces, the distribution of **L** is uniform along the third direction, and the component of **L** along the third direction maintains as zero. Assuming that the wave vector of a spin cycloid is **k**, the distribution of **L** is given by

$$\mathbf{L}=\mathbf{e}_1\cos(\mathbf{k}\cdot\mathbf{x})+\mathbf{e}_2\sin(\mathbf{k}\cdot\mathbf{x}), \quad \text{(S4)}$$

where **x** is a spatial vector in 3D. Since **P** is perpendicular to **k**, and **k** is also within the 2D plane, we have either $\mathbf{k}=|\mathbf{k}|\mathbf{e}_2$, or $\mathbf{k}=-|\mathbf{k}|\mathbf{e}_2$, which correspond to opposite rotation senses with the fixed $\mathbf{e}_2$. The two rotation senses produce different values in the energy term $\gamma\mathbf{P}\cdot[\mathbf{L}(\nabla\cdot\mathbf{L})-(\mathbf{L}\cdot\nabla)\mathbf{L}]$, which is linearly proportional to the gradient of **L**. Therefore, based on the sign of $\gamma$, either the cycloid with $\mathbf{k}=|\mathbf{k}|\mathbf{e}_2$ or $\mathbf{k}=-|\mathbf{k}|\mathbf{e}_2$ has the lower energy. Considering only the low energy state, we have the relation $\mathbf{e}_2=-\frac{\gamma\mathbf{k}}{|\gamma\mathbf{k}|}$. Then Eq. (S4) becomes

$$\mathbf{L}=\frac{\mathbf{P}}{|\mathbf{P}|}\cos(\mathbf{k}\cdot\mathbf{x})-\frac{\gamma\mathbf{k}}{|\gamma\mathbf{k}|}\sin(\mathbf{k}\cdot\mathbf{x}), \quad \text{(S5)}$$

Eq. (S5) can be further modified to

$$\mathbf{L}=-\frac{\gamma\mathbf{P}}{|\gamma\mathbf{P}|}\cos(\mathbf{k}\cdot\mathbf{x})+\frac{\mathbf{k}}{|\mathbf{k}|}\sin(\mathbf{k}\cdot\mathbf{x}), \quad \text{(S6)}$$

When $\gamma>0$, Eqs. (S5) and (S6) are differentiated by a phase shift, i.e., $x\rightarrow x+\pi$. When $\gamma<0$, Eqs. (S5) and (S6) are equivalent. Note that the sign of $\gamma$ is a property of BFO. Eq. (S6) is Eq. (4) in the main text. Eqs. (S5) and (S6) indicate $\mathbf{L}(\mathbf{k})=\mathbf{L}(-\mathbf{k})$. The physical meaning is that if **P** is fixed, and the **k** vector rotates to its opposite direction, then the spins are the same with the original configuration, as shown in Fig. 1(a).

## V. Free energy as a function of the wavelength of the spin cycloid

First we study the free energy as a function of the wavelength based on analytical derivation. To simplify the problem, we only consider the exchange energy and inhomogeneous magnetoelectric interaction. The total free energy density is given by

$$f=A\sum_{i=x,y,z}(\nabla L_i)^2+\gamma\mathbf{P}\cdot[\mathbf{L}(\nabla\cdot\mathbf{L})-(\mathbf{L}\cdot\nabla)\mathbf{L}], \quad \text{(S7)}$$

**L** can be written as $\mathbf{L} = (\sin\theta\cos\phi, \sin\theta\sin\phi, \cos\theta)$ in the spherical coordinate system as established in Fig. S1(a). Polarization **P** is assumed to be along [111]. If the polarization direction is set as the polar axis direction, Eq. (S7) can be rewritten as [6]

$$\begin{aligned} f &= A((\nabla\theta)^2 + \sin^2\theta(\nabla\phi)^2) \\ &+ \gamma(\nabla_x\theta\cos\phi + \nabla_y\theta\sin\phi - \cos\theta\sin\theta(\sin\phi\nabla_x\phi - \cos\phi\nabla_y\phi)) \end{aligned}, \tag{S8}$$

Since the modulation direction $\mathbf{k} = [1\bar{1}0]_{pseudocubic}$ is perpendicular to the polarization direction, we define the **k** vector direction as the *y* axis of a Cartesian coordinate. Since **L** is rotated within the plane spanned by $[111]_{pseudocubic}$ and $[1\bar{1}0]_{pseudocubic}$, we have $\phi = \frac{\pi}{2}, \theta[x, y, z] = \theta[y]$, i.e., $\phi$ is a constant, and $\theta$ is a function of *y*. Eq. (S8) is simplified to

$$f = A\left(\frac{\partial\theta}{\partial y}\right)^2 + \gamma\frac{\partial\theta}{\partial y}, \tag{S9}$$

The Euler-Lagrange equation gives

$$\frac{\delta f}{\delta\theta} = -A\frac{\partial^2\theta}{\partial y^2} = 0, \tag{S10}$$

Therefore, $\theta[y] = \frac{2\pi}{\lambda}y + y_0$, where $\lambda$ is the wavelength, and $y_0$ is the integration constant. If we set $y_0$ as 0, and substitute the solution into Eq. (S9), we have

$$f = A\left(\frac{2\pi}{\lambda}\right)^2 - \gamma\frac{2\pi}{\lambda}, \tag{S11}$$

Eq. (S11) is plotted as a red line in Fig. S1(b), in comparison with the numerical results from phase-field simulations. The analytical and numerical results show good agreement, which both give an energy minimum at equilibrium wavelength $\lambda_0 = \left|-\frac{4\pi A}{\gamma}\right| \sim 62\ \text{nm}$. The value of the equilibrium wavelength is in good agreement with experimental measurements [7].

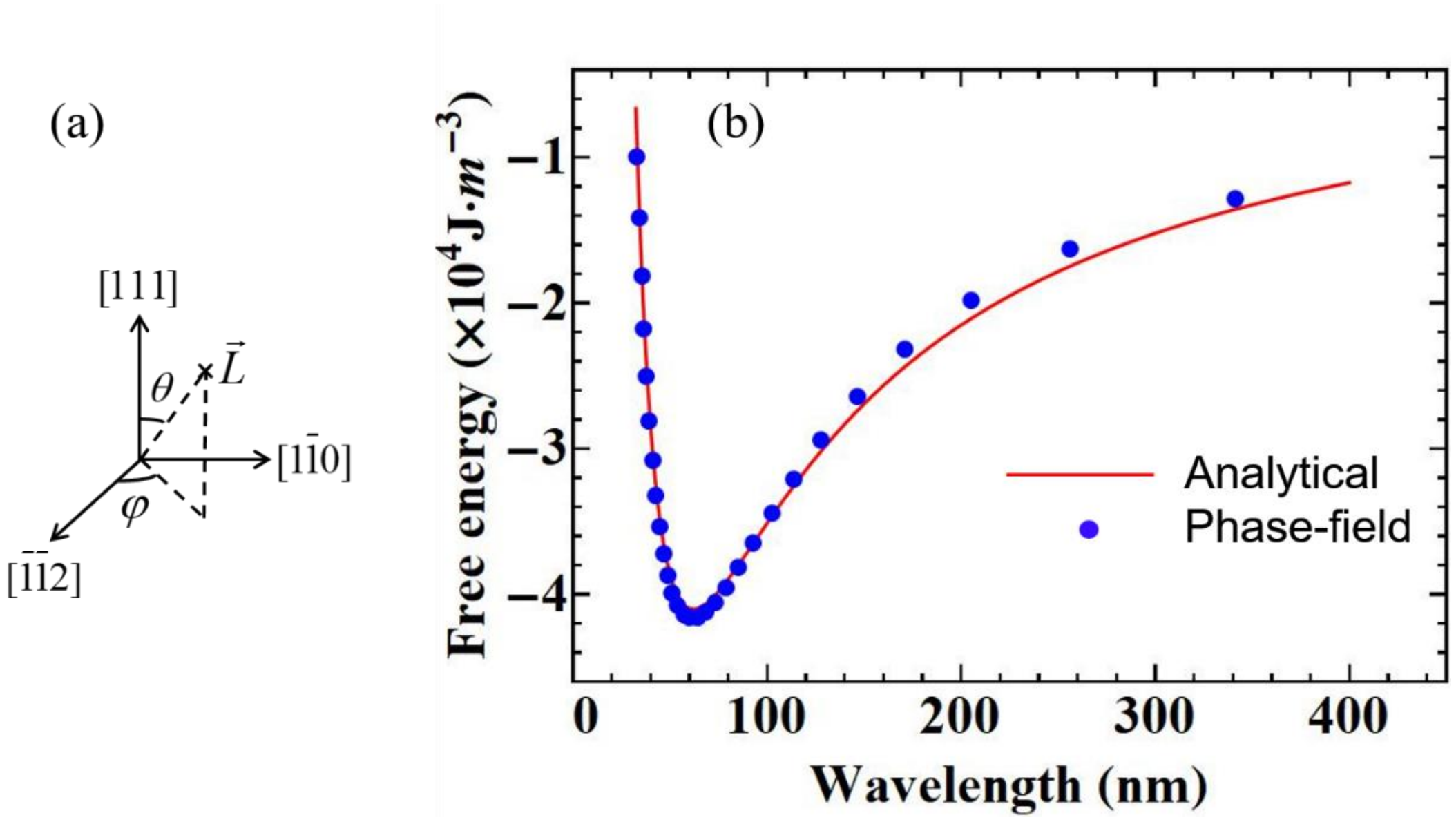

FIG. S1. (a) Schematic of a spherical coordinate system with the polar axis aligned with the $[111]_{\text{pseudocubic}}$ direction and the azimuthal axis along $[\bar{1}\bar{1}2]$. (b) Free energy as a function of the wavelength of the spin cycloid. The red line represents the analytical solution, while the blue dots are from phase-field simulations.

## VI. Free energy as a function of modulation direction of the spin cycloid

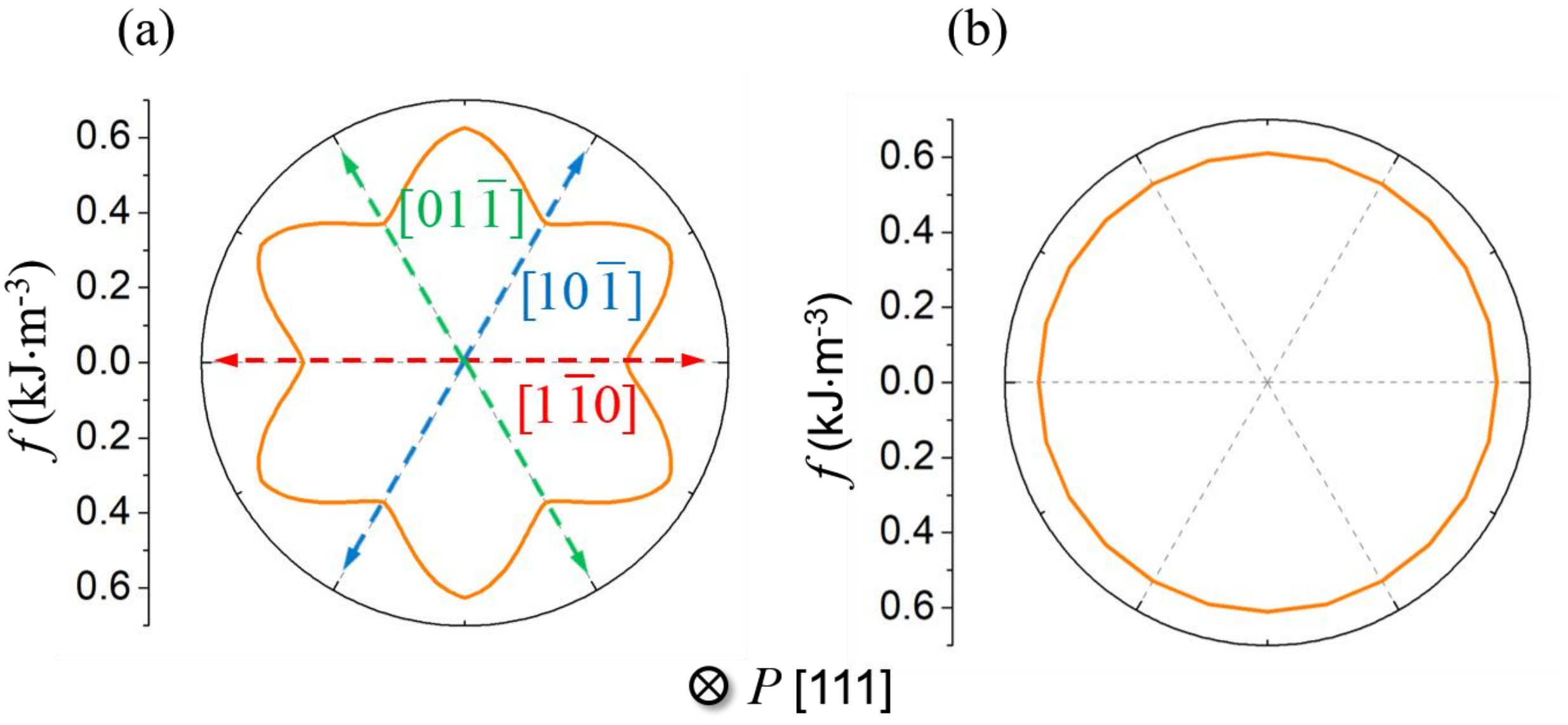

FIG. S2. Polar plot of the free energy versus the direction of the **k** vector for (a) $K_1 = -1.0\times10^4\ \mathrm{J\cdot m^{-3}}$, $K_2 = 3.1\times10^4\ \mathrm{J\cdot m^{-3}}$ , and (b) $K_1 = K_2 = 0.0\ \mathrm{J\cdot m^{-3}}$ . The polarization points along the direction of into the plane of the page.

## VII. System setting for the calculation of domain wall energies

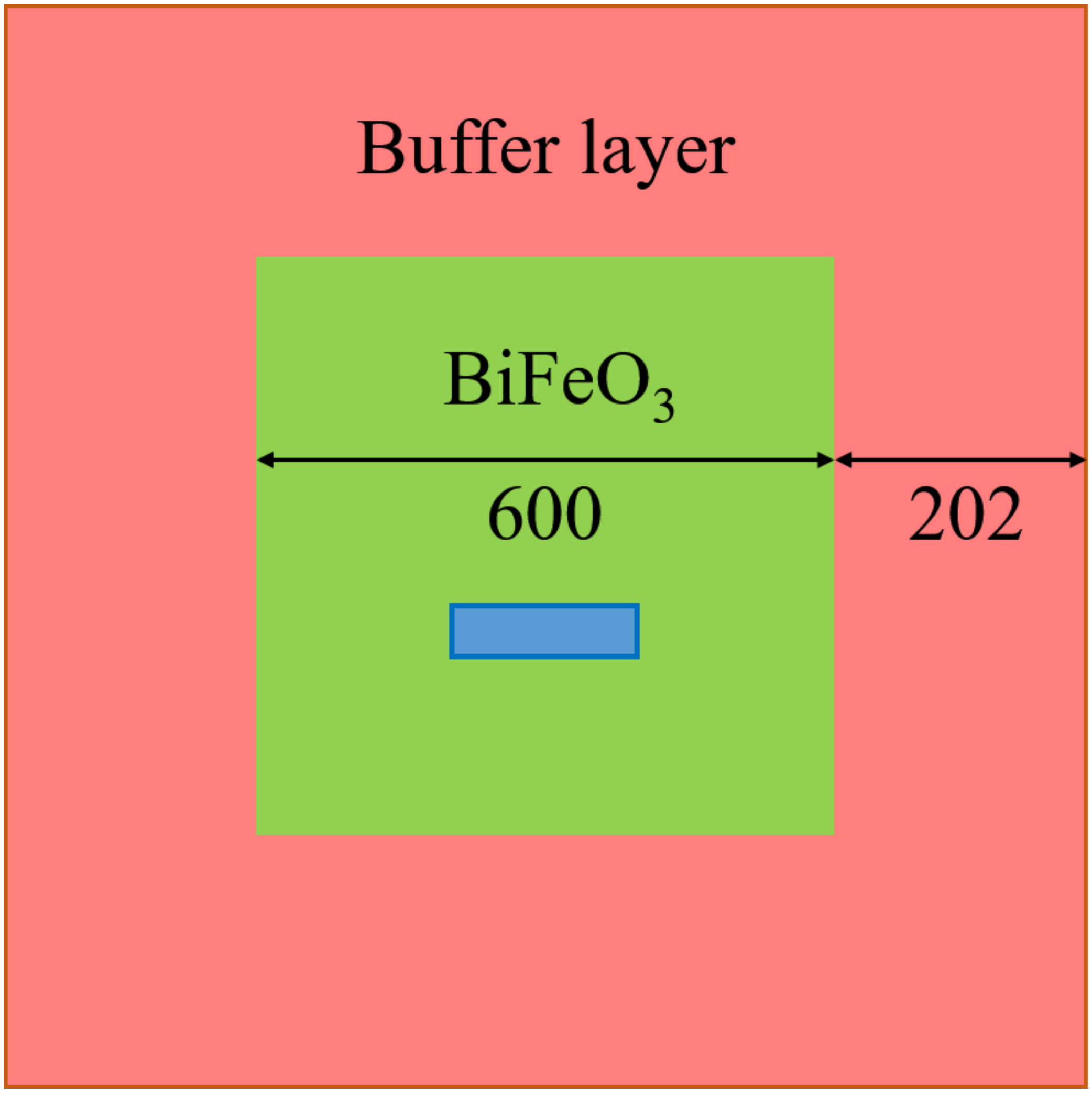

FIG. S3. Schematic for system setting to remove the influence of periodic boundary conditions. Within the green box is $BiFeO_3$, and the surrounding pink regions are buffer layers where **L** is maintained as **0**. The blue box indicates the region within which the spins are plotted in Figs. 2-4.

## VIII. Spin evolution during the (109°, 180°) switching

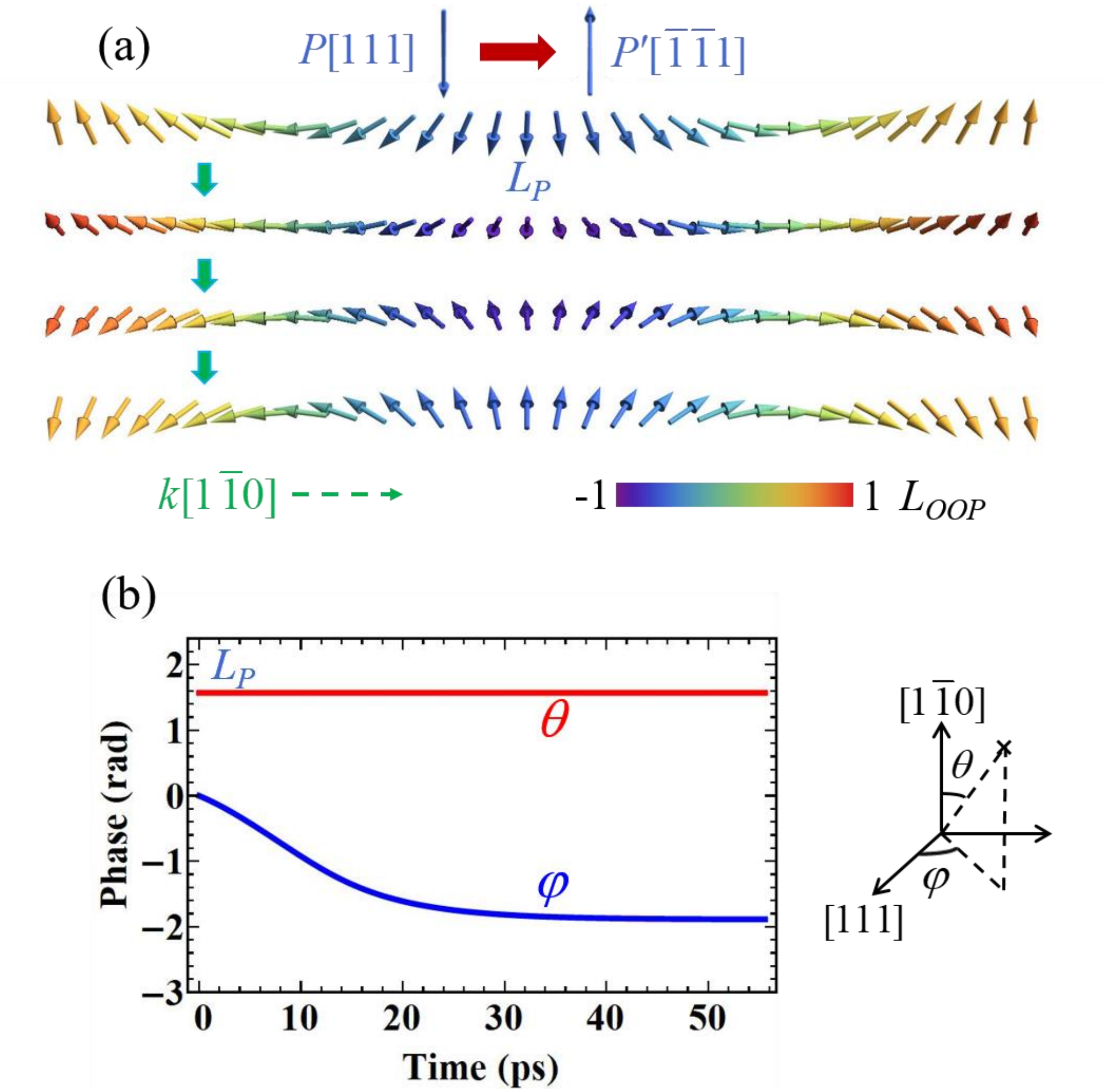

FIG. S4. (a) Four snapshots during the (109°, 180°) switching. The colors of the arrows indicate the component along out-of-the-plane (OOP) of the page. $L_P$ is the spins with the initial **L** parallel to **P**. (b) Evolution of $L_P$ during the (109°, 180°) switching using a spherical coordinate. In the spherical coordinate system setting, the direction of the **k** vector is chosen as the polar axis, and the initial polarization direction is chosen as the reference direction of the azimuth angle $\varphi$.

## IX. Spin evolution during the (71°, 90°) switching

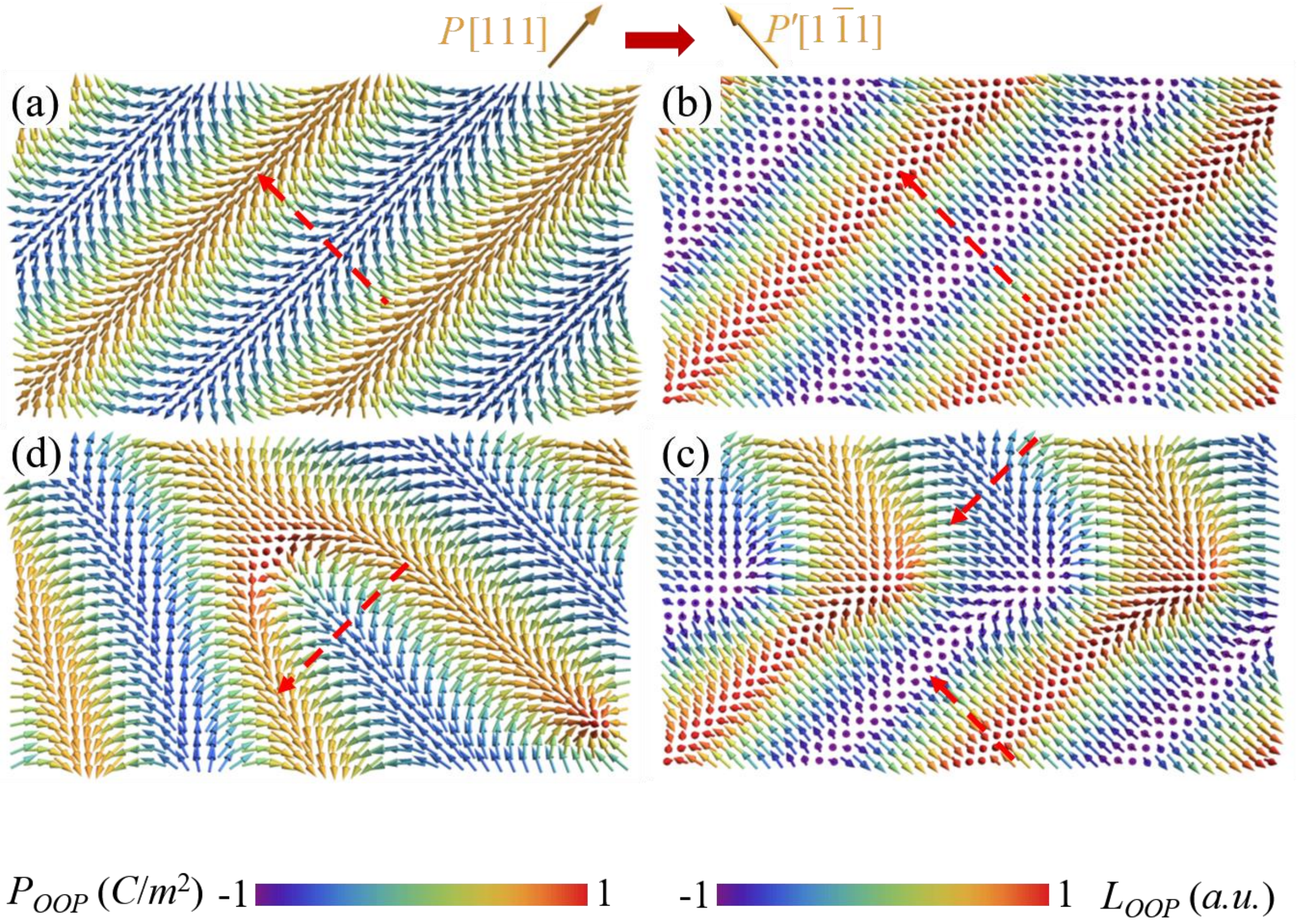

FIG. S5. Spin evolution during the (71°, 90°) switching. Four zoomed snapshots from a 2D simulation, which are the results at (a) 0 ps, (b) 80 ps, (c) 956 ps, and (d) 10,352 ps. The colors of the solid arrows indicate the component along out of the plane of the page. The red dashed arrows specify the **k** direction.

## X. Spin evolution during the (109°, 120°) switching

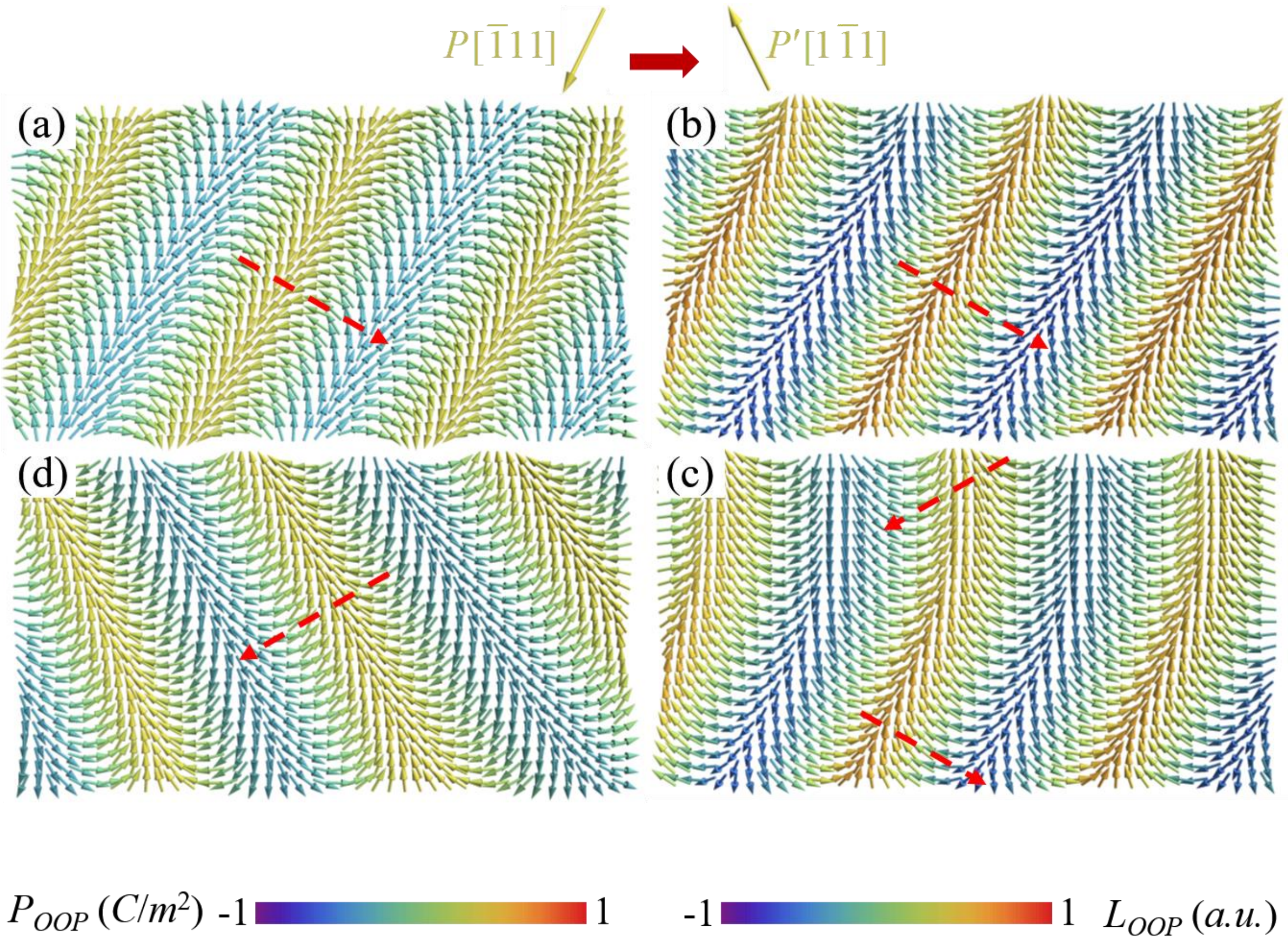

FIG. S6. Spin evolution during the (109°, 120°) switching. Four zoomed snapshots from a 2D simulation, which are the results at (a) 0 ps, (b) 80 ps, (c) 398 ps, and (d) 10,352 ps. The colors of the solid arrows indicate the component along out of the plane of the page. The red dashed arrows specify the **k** direction.

## XI. Spin evolution during the (109°, 90°) switching

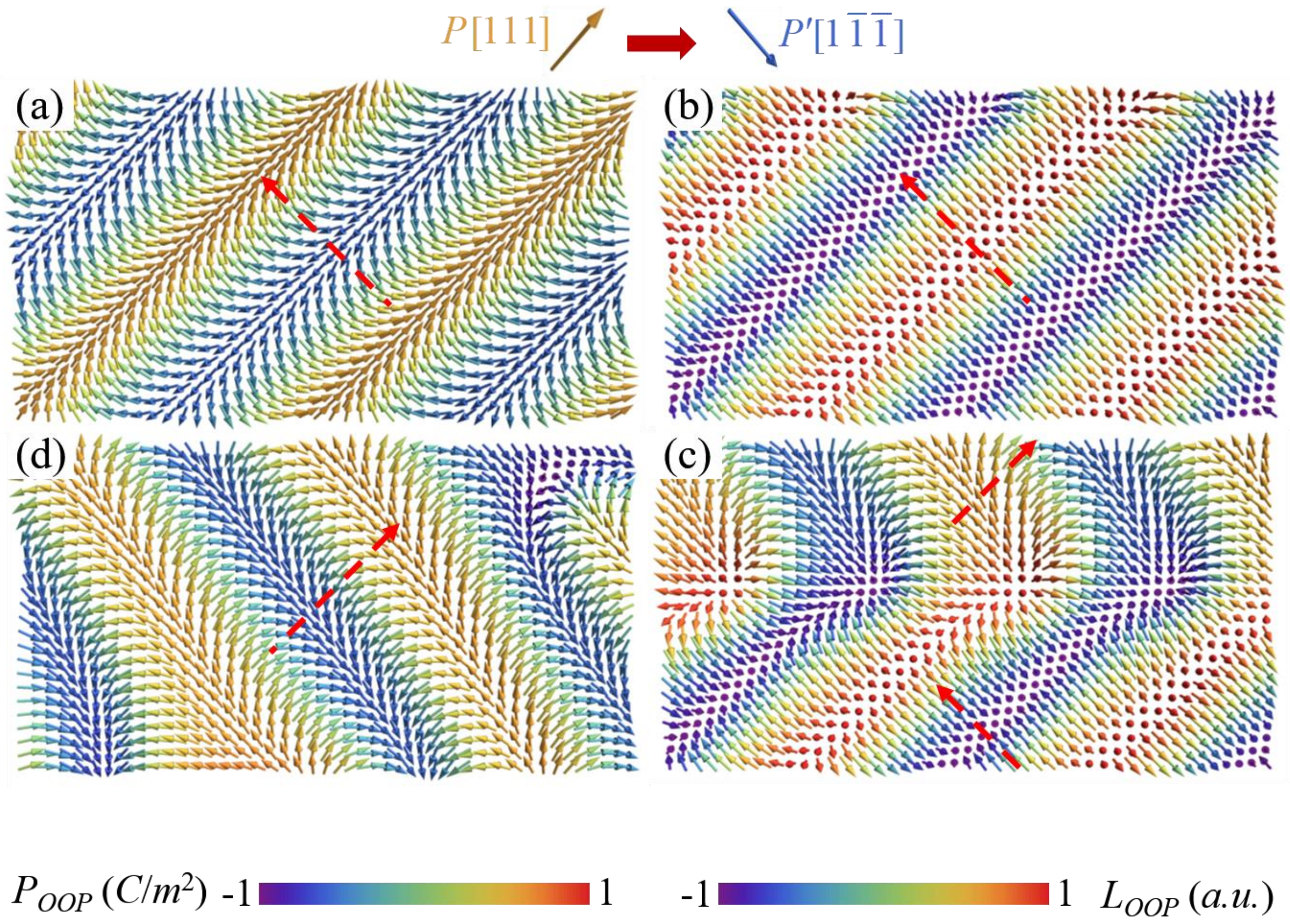

FIG. S7. Spin evolution during the (109°, 90°) switching. Four zoomed snapshots from a 2D simulation, which are the results at (a) 0 ps, (b) 119 ps, (c) 956 ps, and (d) 10,352 ps. The colors of the solid arrows indicate the component along out of the plane of the page. The red dashed arrows specify the **k** direction.

## References

[1] K. M. Hals, Y. Tserkovnyak, and A. Brataas, *Phenomenology of current-induced dynamics in antiferromagnets*, Physical review letters **106**, 107206 (2011).

[2] V. Baltz, A. Manchon, M. Tsoi, T. Moriyama, T. Ono, and Y. Tserkovnyak, *Antiferromagnetic spintronics*, Reviews of Modern Physics **90**, 015005 (2018).

[3] J. Lu, A. Günther, F. Schrettle, F. Mayr, S. Krohns, P. Lunkenheimer, A. Pimenov, V. Travkin, A. Mukhin, and A. Loidl, *On the room temperature multiferroic BiFeO 3: magnetic, dielectric and thermal properties*, The European Physical Journal B **75**, 451 (2010).

[4] I. Sosnowska, W. Schäfer, W. Kockelmann, K. Andersen, and I. Troyanchuk, *Crystal structure and spiral magnetic ordering of BiFeO 3 doped with manganese*, Applied Physics A **74**, s1040 (2002).
[5] J. Zhang and L. Chen, *Phase-field microelasticity theory and micromagnetic simulations of domain structures in giant magnetostrictive materials*, Acta Materialia **53**, 2845 (2005).
[6] D. Sando, A. Agbelele, D. Rahmedov, J. Liu, P. Rovillain, C. Toulouse, I. Infante, A. Pyatakov, S. Fusil, and E. Jacquet, *Crafting the magnonic and spintronic response of BiFeO 3 films by epitaxial strain*, Nature materials **12**, 641 (2013).
[7] I. Sosnowska, T. P. Neumaier, and E. Steichele, *Spiral magnetic ordering in bismuth ferrite*, Journal of Physics C: Solid State Physics **15**, 4835 (1982).